\documentclass[prc,twoside,superscriptaddress,showkeys,showpacs,twocolumn,nofootinbib]{revtex4-1}
\usepackage[pdftex,colorlinks,citecolor=blue,bookmarks]{hyperref}
\usepackage{graphicx, latexsym, amssymb, amsmath, color, multirow, mathrsfs, CJK, ifpdf}
\usepackage[section]{placeins}
\usepackage{amsmath}
\usepackage{amsfonts}
\usepackage{amssymb}
\usepackage{bm}
\usepackage{graphicx}
\usepackage{color} %
\usepackage{txfonts}
\usepackage{float}   
\makeatletter

\newcommand{\Rmnum}[1]{\expandafter\@slowromancap\romannumeral#1@}
\makeatother

\begin{document}

\title{Symmetry Conserving Configuration Mixing description of odd mass nuclei.} 

\author{M. Borrajo}
\author{J. Luis Egido}
\email{j.luis.egido@uam.es}
\affiliation{Departamento de F\'isica Te\'orica, Universidad Aut\'onoma de Madrid, E-28049, Madrid, Spain}

\date{\today}

\begin{abstract}
 We present a self-consistent theory for the description of the spectroscopic properties of odd nuclei which
 includes exact blocking, particle-number and angular-momentum projection and configuration mixing.
 In our theory the pairing correlations are treated in a variation-after-projection approach and the triaxial deformation
 parameters are explicitly considered as generator coordinates. The angular-momentum and particle-number symmetries are exactly recovered. The use of the effective finite-range density-dependent Gogny force in the calculations provides an added value to the theoretical results.
 
 We apply the theory to the textbook example of $^{25}$Mg and, although this nucleus has been thoroughly studied in the
 past, we still provide a novel view of nuclear phenomena taking place in this nucleus. We obtain an overall good agreement
 with the known experimental energies and transition probabilities without any additional parameter such as effective charges.
 In particular, we clearly identify six bands, two of which we interpret  as collective $\gamma$-bands.
  
 \end{abstract}

\pacs{21.60.Jz, 21.10.Dr, 21.10.Ky, 21.10.Re}
\maketitle

\section{Introduction}

The theoretical developments that have taken place in the last years with effective forces  in beyond mean field approaches (BMFA) have allowed to extend the traditional domain of these forces to the full nuclear spectroscopy. The calculations have been performed with the Skyrme  \cite{BHR.03}, the relativistic  \cite{NVR.11} and the Gogny  \cite{EG.16} interactions.

The breakthrough has been possible by the recovery of the symmetries broken in the mean-field approach (MFA)  and by the explicit consideration of large-amplitude fluctuations around the most probable mean-field values. The shape parameters $(\beta,\gamma)$ \cite{Skyrme,RE.10,relativistic} (and pairing gaps \cite{Nuria,Nuria2,Nuria3}) have been  used as coordinates in the framework of the generator-coordinate method (GCM) and the particle-number (PN) and angular-momentum (AM) symmetries  recovered by means of projectors. The most sophisticated level has been reached by considering the cranking frequency as an additional generator coordinate \cite{BRE.15,EBR.16,EG.16},
which considerably improves the results and allows the study of new phenomena.
These developments are called symmetry-conserving configuration mixing  (SCCM) approaches and  so far  have only been applied to even-even nuclei. Methods based on the Bohr collective Hamiltonian have also made significant progress lately \cite{Gogny_B, Skyrme_B, relativistic_B}.   

Calculations for odd-even and odd-odd nuclei are not as much  developed as those for even-even ones. The reason is  that odd nuclei  are far more complicated to deal with.  Already at the mean field,   in the   BCS approach or in Hartree-Fock-Bogoliubov (HFB) theories,  they are numerically awkward and one must consider several channels (spins, parity, etc) to find the ground state.   An additional difficulty is the breaking of the time-reversal symmetry by the blocked structure of the wave  function and the fact that triaxial calculations must be performed.    In spite of these difficulties  it seems natural to extend the above-mentioned approaches to odd-even and odd-odd nuclei. As a matter of fact angular-momentum projected calculations  for odd-A nuclei started long ago, though they have been mostly performed on HF or HFB states  in small valence spaces \cite{BGR.65,GW.67,RPK.93,HI.84,HSG.85}. More recently, a GCM mixing based on parity and AM-projected  Slater determinants in a model space of antisymmetrized Gaussian wave packets has been carried out in the frameworks of fermionic \cite{NF.08} and antisymmetrized \cite{KTK.13,KK.10}  molecular dynamics. In the latter calculations, however,  the pairing correlations are not treated properly.  A preliminary  BMFA study of odd-even nuclei with the Skyrme force has been presented in Ref.~\cite{Bally}. Our first BMFA applications to odd-nuclei  with the Gogny force did not consider configuration mixing. Thus,  in Ref.~\cite{BE.16} the  nucleus $^{31}$Mg at the border of the $N=20$ inversion island was studied, with relevant contributions to the understanding of the shape coexistence phenomenon in excited states. More recently, an exhaustive study of the ground state properties in the magnesium isotopic chain 
with the Gogny force has been performed in Ref.\cite{BE.17}.  Excellent agreement has been obtained for binding energies, one-neutron separation energies, odd-even mass differences, radii, quadrupole and magnetic moments, etc.
 
 In this work we generalize the full SCCM approach with the Gogny interaction to the description of spectroscopic properties of odd-even nuclei.  Specifically, we consider linear combinations of PN and AM projected, exactly blocked,  triaxial HFB wave functions generated in the $(\beta,\gamma)$ plane.
  As an application we have chosen the nucleus $^{25}$Mg, which has widely been studied theoretically and experimentally in the past.
 The reason for this choice is that this nucleus presents collective as well as single particle degrees of freedom. Furthermore,  the knowledge of many experimental properties will allow us to make a thorough  check of our theory.
 
 In Sect.~\ref{Sect:Theory} we outline  the theoretical methods used in the calculations. In Sect.~\ref{Sect:spe_PES} we 
 present the single particle aspects and the  potential energy surfaces. In Sect.~\ref{Sect:SCCM},  the SCCM results are discussed with special emphasis on the different bands and transition probabilities. We finish this work with a summary and the corresponding conclusions. 
   
\section{Theory}\label{Sect:Theory}
As mentioned in the Introduction, our SCCM wave functions are written  as a linear combination of PN and AM projected blocked HFB wave functions generated with the quadrupole moments as coordinates.
In this section, in a first step we explain how the HFB wave functions are generated and in the following we describe the way in which  the SCCM equations are solved.

\subsection{The blocked equations} \label{susect:bl_eq}
The cornerstone of the  BMFA  is the HFB theory \cite{RS.80}. The  HFB  wave function $|\phi\rangle$ is a product of  quasiparticles $\alpha_{\rho}$ defined by  the  transformation
\begin{equation}  \label{bogtrans}
\alpha^{\dagger} _\rho=\sum_{\mu=1}^{2M} U_{\mu \rho}^{}c_{\mu}^{\dagger}+V_{\mu \rho}^{}c_{\mu},
\end{equation}
where ${c_{\mu}^{\dagger},c_{\mu}}$ are the particle-creation and -annihilation operators in the reference  basis, in our case the Harmonic Oscillator one. The matrices  $U$ and $V$ are determined by the variational principle.   

As usual  we impose three discrete self-consistent symmetries on our basis states $\{c_{\mu}^{\dagger},c_{\mu}\}$: spatial parity, $\hat{P}$, simplex,  $\Pi_{1}=\hat{P}e^{-i\pi J_{x}}$ and the  $\Pi_{2} {\cal T}$ symmetry, with $\Pi_{2}= \hat{P}e^{-i\pi J_{y}}$ and  ${\cal T}$ the time-reversal operator. 
The first two symmetries provide good parity and simplex quantum numbers and the third allows to use only real quantities. The simplex symmetry furthermore characterizes the blocking structure of odd and even nuclei \cite{Ma.75,EMR.80,BE.17}.  The single particle basis states are symmetrized  in such a way that
\begin{equation}  \label{simpsim}
\Pi_{1} c^{\dagger}_k  \Pi_{1}^{\dagger}  = +i  c^{\dagger}_k,   \;\;\;
\Pi_{1} c^{\dagger}_{\overline k}  \Pi_{1}^{\dagger}  = -i  c^{\dagger}_{\overline k}.
\end{equation}
with ${ k=1,...,M}$ and $2M$ the dimension of the configuration space. We use latin indices  to distinguish the levels according to their simplex, $\{k,l,m\}$ for simplex $+i$ and $\{{\overline k},{\overline l},{\overline  m}\}$ for simplex $-i$. Greek indices on the other hand  do not distinguish simplex and run over the full configuration space. Notice furthermore that with our single particle symmetrisation the states $c^{\dagger}_k$ and  $c^{\dagger}_{\overline k}$ are related by time reversal symmetry, i.e., 
${\cal T}c^{\dagger}_k {\cal T}^{\dagger}= c^{\dagger}_{\overline k} $.

If we impose the intrinsic wave function  $|\phi\rangle$ to be an eigenstate of the simplex operator, then, for a paired even-even nucleus,  half of the quasiparticle operators $\alpha^{\dagger}_{\mu}$  have simplex $+i$ and the other half 
have simplex $-i$, i.e., Eq.~(\ref{bogtrans}) separates in two blocks~:
\begin{eqnarray}  \label{bogtrans_simp} 
\alpha _m^{\dagger} &= &\sum_{k=1}^{M}U_{km}^{+}c_k^{\dagger} +V_{km}^{+}c_{\overline k}^{}, \nonumber \\
\alpha _{\overline m}^{\dagger} &= &\sum_{k=1}^{M}U_{km}^{-}c_{\overline k}^{\dagger} +V_{km}^{-}c_k^{}, 
\end{eqnarray}
with ${ m=1,...,M}$ in an obvious notation. 

The  wave function of the ground state of an even-even nucleus is given by 
\begin{equation}
| \phi \rangle = \prod_{\mu=1}^{2M} \alpha_\mu  |-\rangle,
\label{Eq:vac}
\end{equation}
with $|-\rangle$ the particle vacuum\footnote{In the product only  quasiparticle operators that do not annihilate trivially the particle vacuum are allowed.}.  The quasiparticle vacuum $|\phi\rangle$ is obviously defined by
\begin{equation}
\alpha_{\mu}  | \phi \rangle = 0, \;\;   \mu=1,...,2M.
\label{Eq:qp_vac}
\end{equation}
Since in Eq.~\ref{Eq:vac}  there are as many quasiparticle operators with simplex $+i$ as with $-i$, the ground state of an even-even nucleus has simplex $+1$. The quasiparticle excitations 
\begin{equation}
| {\tilde \phi}^{\pi} \rangle =  \alpha^{\dagger}_{\rho_1} |\phi \rangle
\label{eq:odd_even_ansatz}
\end{equation}
correspond to  odd-even nuclei. They can be written as  vacuum
to the quasiparticle operators ${\tilde \alpha}_\rho$, 
\begin{equation}
{\tilde \alpha}^{}_\rho| {\tilde \phi}^{\pi} \rangle =0, \;\;   \rho=1,...,2M.
\label{Eq:vacexc}
\end{equation}
The $2M$ operators 
\begin{equation}  \label{bogtransb}
	{\tilde \alpha}^{\dagger} _\rho=\sum_{\mu}{\tilde U}_{\mu \rho}^{}c_{\mu}^{\dagger}+{\tilde V}_{\mu \rho}^{}c_{\mu},
\end{equation}
 are obtained from the set $\{ { \alpha}^{\dagger}_\mu \}$ by replacing the creation operator $\alpha^{\dagger}_{\rho_1}$ by the annihilation operator $\alpha^{}_{\rho_1}$, the other $2M-1$  operators remain unchanged. The simplex of the state $|{\tilde \phi}^{\pi} \rangle$ is given by
  \begin{equation}
  \Pi_1| {\tilde \phi}^{\pi} \rangle = i^{n}| {\tilde \phi}^{\pi} \rangle
 \end{equation} 
  where we have introduced the blocking number $n$. It is $n=1$ if $\alpha^{\dagger}_{\rho_1}$ has simplex $+i$ and
$n=-1$ if   $\alpha^{\dagger}_{\rho_1}$ has simplex $-i$. The symbol $\pi$ indicates the parity of the state $|{\tilde \phi}^{\pi}\rangle$.
    Notice that in the running product of Eq.~(\ref{Eq:vac}), orbitals with the same parity are occupied pairwise.  Therefore, the parity $\pi$ of the state  $|{\tilde \phi}^{\pi} \rangle$ is given by the parity of the blocked level $\alpha^{\dagger}_{\rho_{1}}$.
    In this work we are interested in $^{25}$Mg. Since the magnesium isotopes have $Z=12$, we restrict ourselves to the neutron channels. We therefore consider  wave functions of the form of Eq.~(\ref{eq:odd_even_ansatz})
where  $\rho_{1}$ denotes a neutron state.
According to the  parity we have two blocking channels:  neutrons of positive or negative parity. 
Once the isospin and the parity are chosen one must furthermore decide the simplex of the state to block, i.e., $+i$ or $-i$.
    However, if the HFB Hamiltonian is time reversal invariant, the matrices $\{\tilde{U},\tilde{V}\}$ of the Bogoliubov transformation obtained either from the solution of the HFB equations  or from  Eq.~(\ref{E_Lagr_bet-gam}) are such that  ${\cal T}\alpha^{\dagger}_k {\cal T}^{\dagger}= \alpha^{\dagger}_{\overline k} $. In this case the HFB states  $ \alpha^{\dagger}_{k} |\phi \rangle$ and $\alpha^{\dagger}_{\overline k} |\phi \rangle$ obtained by blocking a positive and a negative simplex state, respectively, see Eq.~(\ref{eq:odd_even_ansatz}), are related by the  time reversal symmetry, $\alpha^{\dagger}_{\overline k} |\phi \rangle= {\cal T}\alpha^{\dagger}_{k} |\phi \rangle$, and  are degenerated (Kramers degeneracy).  Since this is our case, see Eq.~(\ref{E_Lagr_bet-gam}) below,  we only need to block a quasiparticle with a given simplex\footnote{This argument is correct for a general Hamiltonian. For a density dependent interaction it works also but the demonstration is a bit more elaborated.}. Notice that in the case of the cranking Hamiltonian, $\hat{H}^{\prime} = \hat{H} - \omega \hat{J}_{x}$, the former statement is not  correct.
    
  Though the state $|{\tilde \phi}^{\pi} \rangle$ has the right blocking structure, $|{\tilde \phi}^{\pi} \rangle$ is not an eigenstate of the PN   or the AM operators since the Bogoliubov transformation mixes creator and annihilator operators and states with different angular momenta.  As for even-even nuclei, to recover the particle-number symmetry one has to project to the right quantum numbers, see \cite{RS.80}. The {\em easiest} way to recover the symmetries would be to minimize the HFB energy, i.e., determine $({\tilde U},{\tilde V})$ and then perform the projections. This is the so-called projection-after-variation (PAV) approach. The {\em optimal}  way is to determine $({\tilde U},{\tilde V})$ directly from the minimisation  of the projected energy, i.e, the variation-after-projection (VAP) method. From even-even nuclei one knows that PN-VAP is feasible  while AM-VAP is  very
CPU-time consuming.  The approach of solving  the PN-VAP variational equation to find the self-consistent minimum and afterwards to perform an  AM-PAV is not very good because the AMP is not able to exploit any degree of freedom of the HFB transformation and self-consistency with respect to the AMP is therefore not guaranteed.

  An interesting  option  is to perform an approximate AM-VAP approach as  it  has been used in the projected mean field  theory of Refs.~\cite{BE.16,BE.17}. In this approach the variational PN-VAP equation is solved for a large set of relevant physical situations (wave functions) as to cover the sensitive degrees of freedom to  the AM projection.  Afterwards for each angular momentum one calculates the  AM-PAV energy with this set of wave functions to determine the absolute minimum among these states. This procedure provides  different HFB wave functions for unlike AM. In Refs.~\cite{BE.16,BE.17} the deformation parameters $(\beta,\gamma)$ were considered as the additional degrees of freedom since they are believed to provide the strongest energy dependence of the nuclear interaction with the AM.  This method guarantees, at least, AM-VAP self-consistency with respect to these relevant  quantities.  Notice that we obtain  approximate AM-VAP solutions for the projected mean field theory at the cost of performing AM-PAV in the  $(\beta,\gamma)$ grid for each angular momentum, see Figs.~\ref{Fig:PES_POS},\ref{Fig:PES_NEG} below.  Though in this work we are not performing projected mean field calculations we will see in the next subsection that  this feature has consequences for the SCCM calculations of this work.
 	
 	As mentioned above the SCCM aims to describe vibrations associated to the shape parameters and towards this end a superposition of  wave functions with different $(\beta,\gamma)$ is considered, see Eq. \ref{GCM_ANSA} below.	In order to generate the wave functions we solve the PN-VAP  constrained equations on a grid of $(\beta,\gamma)$  points:
\begin{eqnarray}
{E^{\prime}}[{\tilde \phi}^{\pi}]= \frac{ \langle{\tilde \phi}^{\pi}|\hat{H}\hat{P}^{N}|{\tilde \phi}^{\pi} \rangle}{\langle{\tilde \phi}^{\pi}|\hat{P}^{N}|{\tilde \phi}^{\pi} \rangle} -  \langle {\tilde \phi}^{\pi} |\lambda_{q_{0}}\hat{Q}_{20} + \lambda_{q_{2}}  \hat{Q}_{22} | {\tilde \phi}^{\pi} \rangle, \label{E_Lagr_bet-gam}
\end{eqnarray}
with  the Lagrange multiplier $\lambda_{q_{0}}$  and $\lambda_{q_{2}}$ being determined by the constraints 
\begin{equation}
\langle {\tilde \phi}^{\pi} |\hat{Q}_{20} | {\tilde \phi}^{\pi} \rangle =q_{0}, \;\; \; \langle {\tilde \phi}^{\pi} |\hat{Q}_{22} | {\tilde \phi}^{\pi} \rangle =q_{2}. \label{q0_q2_constr}
\end{equation}
The relation between $(\beta,\gamma)$ and $(q_{0},q_{2})$ is given by 
$\beta= \sqrt{20\pi(q_{0}^{2} + 2 q_{2}^{2})}/3r^{2}_{0}A^{5/3}$,
$\gamma = \arctan (\sqrt{2}{q_2}/q_{0})$
with $r_{0}=1.2$ fm and  $A$ the mass number. The solution of Eqs.~(\ref{E_Lagr_bet-gam},\ref{q0_q2_constr}) for a large number of $(\beta,\gamma)$ points determines the set of states  $\tilde{\phi}^{\pi}(\beta,\gamma)$ needed for the calculations.

The minimization of Eqs.~(\ref{E_Lagr_bet-gam}-\ref{q0_q2_constr}) is performed with the conjugated-gradient method \cite{grad}. The blocking structure of the wave function of Eq.~(\ref{eq:odd_even_ansatz}) is a self-consistent symmetry and for a given blocking number we determine the lowest solution in the blocked channel compatible with the imposed constraints.  Therefore, independently of which quasiparticle state of the given  isospin-parity-simplex  channel was initially blocked, at the end of the iteration process we always obtain the same solution.

\subsection{The SCCM method}

The next step is the simultaneous particle-number and angular-momentum projection (PNAMP) of each state   $|{\tilde\phi}^{\pi}(\beta,\gamma)\rangle$  that conforms the $(\beta,\gamma)$ grid. The resulting states are given by
\begin{equation}
 |IMK,\pi,N, (\beta,\gamma) \rangle= P^N P^I_{MK} \; |{\tilde\phi}^{\pi} (\beta,\gamma)\rangle.
  \label{GCM_BASIS}
\end{equation} 
The final SCCM solution we are looking for is given by  
\begin{equation} \label{GCM_ANSA}
|\Psi^{N,I,\pi}_{M,\sigma }  \rangle 
=  \sum_{K,\beta,\gamma} f^{I}_{K\sigma}(\beta,\gamma)  |IMK,\pi,N, (\beta,\gamma) \rangle,  
\end{equation}
where $\sigma$ labels the states with the same quantum numbers and different energies and the  coefficients  $ f^{I}_{K\sigma}(\beta,\gamma)$  are variational parameters.   They are determined by the energy minimization which  provides the Hill-Wheeler-Griffin (HWG) \cite{HWG} equation 
\begin{equation}
\sum_{K'\beta'\gamma'} \, \,(\mathcal{H}^{N,I,\pi}_{\beta \gamma K,\beta' \gamma' K'} - E^{N,I,\pi}_\sigma \mathcal{N}^{N,I,\pi}_{\beta \gamma K,\beta' \gamma' K'}) f^{I}_{K\sigma}(\beta',\gamma')  = 0.
\label{HW_Eq}
\end{equation} 
where $\mathcal{H}^{N,I,\pi}_{\beta \gamma K,\beta' \gamma' K'}$ and $\mathcal{N}^{N,I,\pi}_{\beta \gamma K,\beta' \gamma' K'}$ 
are the Hamiltonian and norm overlaps defined by 
\begin{eqnarray}
\hspace{-0.7cm}\mathcal{H}^{N,I,\pi}_{\beta \gamma K,\beta' \gamma' K'} \!  =  \!   \langle IMK,\pi,N, (\beta,\gamma) |H | IMK',\pi,N,(\beta,\gamma) \rangle \label{hamove} \nonumber \\
\hspace{-0.7cm}{\mathcal{N}^{N,I,\pi}_{\beta \gamma K,\beta' \gamma' K'} \!  =  \!   \langle IMK,\pi,N,(\beta,\gamma) | }IMK',\pi,N,(\beta,\gamma) \rangle \label{normove}.
\end{eqnarray}
The presence of the norm matrix in Eq.~(\ref{HW_Eq}) is due to the non-orthogonality  of the states $|IMK,\pi,N,(\beta,\gamma) \rangle$.

We have seen in the precedent subsection that  considering the $(\beta,\gamma)$ degrees of freedom within the framework of the projected mean field approach, i.e., statically, was equivalent to an approximate AM-VAP at the mean field level. In the SCCM  approach, Eq.~(\ref{GCM_ANSA}),  one performs AM-VAP with respect to the mixing amplitudes, i.e., statical and  dynamical correlations are considered. It seems, therefore, that the AM projection is to a very good approximation a full AM-VAP (with respect to the $(\beta,\gamma)$ degrees of freedom)  at all levels of the calculations.

To solve the HWG equations  one first introduces an orthonormal basis defined by the eigenvalues, $n_{\kappa} ^{I}$, and eigenvectors, $u^{IK}_{\kappa}(\beta,\gamma)$, of the norm overlap:
\begin{equation}
\sum_{\beta'\gamma'K'}\mathcal{N}_{\beta \gamma K,\beta' \gamma' K'}^{NI\pi}  u_{\kappa} ^{IK'}(\beta'\gamma')=n_{\kappa}^{I}u_{\kappa}^{IK}(\beta,\gamma).
\end{equation}
This orthonormal basis is known as the natural basis and,  for $n_{\kappa}^{I}$ values such that  
$n_{\kappa}^{I}/n^{I}_{max}>\zeta$, the  natural states are defined by:
\begin{equation} \label{nat_sta}
|\kappa^{I}\rangle=\sum_{\beta\gamma K}\frac{u^{IK}_{\kappa}(\beta,\gamma)}{\sqrt{n^{\kappa I}}}| IMK,N,(\beta,\gamma)\rangle.
\end{equation}
Obviously, a cutoff  $\zeta$ has to be introduced in the value  of the norm eigenvalues to avoid linear dependences \cite{RingAMP_Rel_09}.
Then, the HWG equation is transformed into a normal eigenvalue problem:
\begin{equation}\label{HWG_nat}
\sum_{\kappa'}\langle\kappa^{I}|\hat{H}|\kappa'^{I}\rangle g^{\sigma I}_{\kappa'}=E^{\sigma I}g^{\sigma I}_{\kappa}.
\end{equation}

In the natural basis the wave function of Eq.~(\ref{GCM_ANSA}) is given by
\begin{eqnarray}
|\Psi^{N,I,\pi}_{M,\sigma }  \rangle =   \sum_{\kappa} g^{\sigma I}_{\kappa} | \kappa ^{I}\rangle.  
 \label{Psi}
\end{eqnarray}
From the coefficients $g^{\sigma I}_{\kappa}$ we can define the quantities 
\begin{equation}
p^{\sigma I}_{K}(\beta,\gamma)=\sum_{\kappa }g^{\sigma I}_{\kappa}u^{IK}_{\kappa}(\beta,\gamma)
\label{coll_wf}
\end{equation} 
that satisfy 
\begin{equation}
\sum_{\beta\gamma K} |p^{\sigma I}_{K}(\beta,\gamma)|^2=1,  \;\;\; \forall \sigma,
\label{norm_coll_wf}
\end{equation}
and are equivalent to a probability amplitude. In terms of these quantities  we can define  the collective wave function
\begin{equation}
\mathcal{P}^{\sigma I}(\beta,\gamma)=\sum_{{K}}\vert p_{K}^{\sigma I}(\beta,\gamma)\vert^{2},
\label{p2}
\end{equation}
which gives  the probability of finding the fixed deformation parameters $(\beta,\gamma)$ for a given $I$ in the $(\beta,\gamma)$ plane.
The collective wave function allows to calculate the average values of different  observables.  
The probability distribution of finding the  projection $K$ in the collective wave function is obtained by summing over all possible deformations:
\begin{equation}
\mathcal{P}^{\sigma I}_{K}=\sum_{{\beta,\gamma}}\vert p_{K}^{\sigma I}(\beta,\gamma)\vert^{2}.
\label{p1}
\end{equation}

The electromagnetic transition probabilities and the
spectroscopic multipole moments for odd-A nuclei are calculated with the same expression as used for the even-even ones, see Ref.~\cite{RE.10,EG.16}.

\section{Results: Single particle energies and potential energy surfaces}\label{Sect:spe_PES}
As an application of our theory we chose the nucleus $^{25}$Mg which has been widely studied both experimentally \cite{Li.58,SS.68,Headly.88,Hei.91} and theoretically \cite{Cole.75,Bally}  and used as prime example in 
several textbooks \cite{BM.75,EG.75}

\begin{figure}[t]
	\begin{center}
		\includegraphics[angle=0,scale=.45]{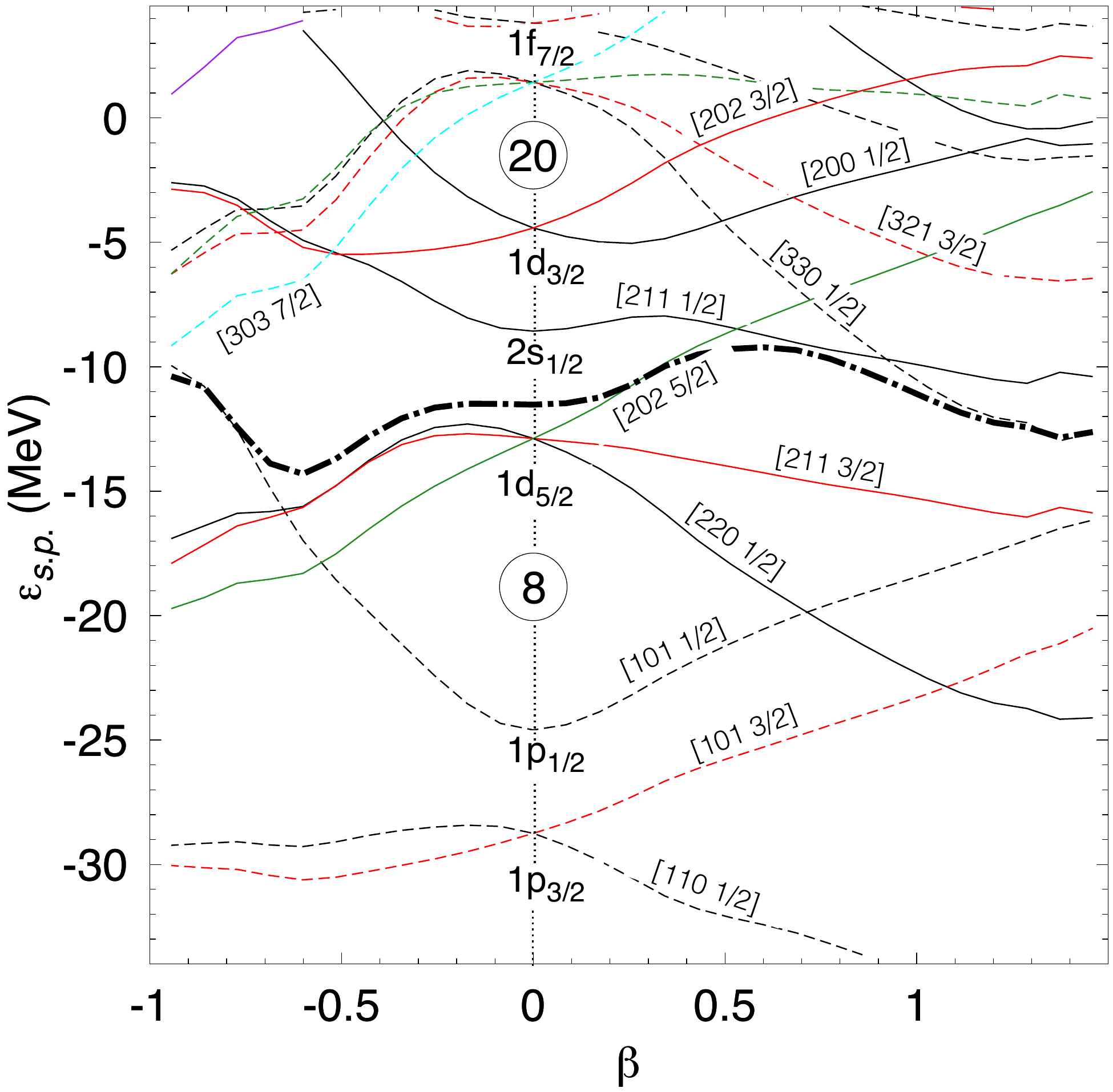}
		\caption{Single-particle levels of $^{25}$Mg for neutrons in the HFB approach. The thick dashed line represents the Fermi level. The Nilsson quantum numbers $[N,n_{z},m_{l},\Omega]$ are indicated for the relevant orbitals.}
		\label{Fig:SPE}       
	\end{center}
\end{figure}

 In the calculations the intrinsic many body wave functions $|{\tilde \phi}^{\pi}(\beta,\gamma)\rangle$ are expanded in a Cartesian harmonic oscillator basis and the number of spherical shells included in this basis is $N_{shells}=8$ with an oscillator length of $b=1.01A^{1/6}$. 
 The $(\beta,\gamma)$ grid spans the  sextant $0 ^{\circ}\leq \gamma \leq 60^{\circ}$ in the range $\beta\leq 1.1 $ $(\beta\leq 1.5) $  and contains  190(216) points for positive (negative) parity.
 
  The  angular momentum projection has been done with the set of integration points in the Euler angles  $(N_{\boldsymbol{\alpha}}=N_{\boldsymbol{\beta}}=N_{\boldsymbol{\gamma}}=32)$ in the intervals $\boldsymbol{ \alpha} \in [0,2\pi ], \boldsymbol{ \beta} \in [0,\pi],\boldsymbol{\gamma} \in [0,2\pi]$. The number of points to perform the integral of the particle-number projection is  $11$. 
In the calculations we use the Gogny interaction \cite{DG.80} with the D1S parameterization \cite{BGG.91}.
We consider all exchange terms of the interaction, the Coulomb force and the two-body correction of the kinetic energy to avoid problems with the PNP \cite{AER.01Ex,AER.01PNP}. Concerning the density dependence of the force we adopt  the projected density prescription for the PNP and the mixed one for the AMP. For further details see for example Refs.~\cite{RE.10,EG.16}.  

\subsection{Single particle levels}

 The neutron single particle energy (spe) levels around the Fermi level play a relevant role in the determination of the  blocked structure of the wave function,  Eq.~(\ref{eq:odd_even_ansatz}),  of an odd nucleus.  Since the blocking breaks the axial symmetry,  in order to produce an ordinary Nilsson plot we have solved the axially symmetric HFB equations without blocking but with the constraint on the number of neutrons $\langle \hat{N} \rangle=13$.
 These energy levels  are given by the solution of the HFB equation for different $\beta$-values and are shown  in Fig.~\ref{Fig:SPE}  for neutrons. The proton single particle  energies for this light nucleus look similar to the neutron ones. This plot allows to guess the quantum number of the lowest blocked state as a function of the deformation.  According to this plot and for positive parity states  the 
 candidates to host the odd neutron for prolate shapes are the [202 5/2] orbital for  $\beta \le 0.52$ and the [211 1/2] for larger $\beta$-values. For oblate shapes the lowest orbitals for the blocked neutron correspond to  the level [220 1/2] for small deformations and the [211 3/2] for larger ones.
 
  In the negative parity channel there are two ways to produce excited states of negative parity: making a hole in the [101 1/2] orbital (1p$_{1/2}$ subshell) or promoting a particle to the [303 1/2] orbital for prolate ([303 7/2] for oblate) shapes (1f$_{7/2}$  subshell). The  [101 1/2] orbital gets close to the Fermi level at $\beta\approx -0.7$ and the orbital [303 7/2] crosses the Fermi level at very large deformations, $\beta \approx -0.9$.  The Nilsson scheme is thought for orientation purposes and the quoted   $\beta$-values are only approximate since, as mentioned above,  the blocking effect has not been taken into account in this plot and our results are based on exact blocking and on the PNVAP approach of  Eq.~(\ref{E_Lagr_bet-gam}).
   \begin{figure}[t]
 	\begin{center}
 		\includegraphics[angle=0,scale=.27]{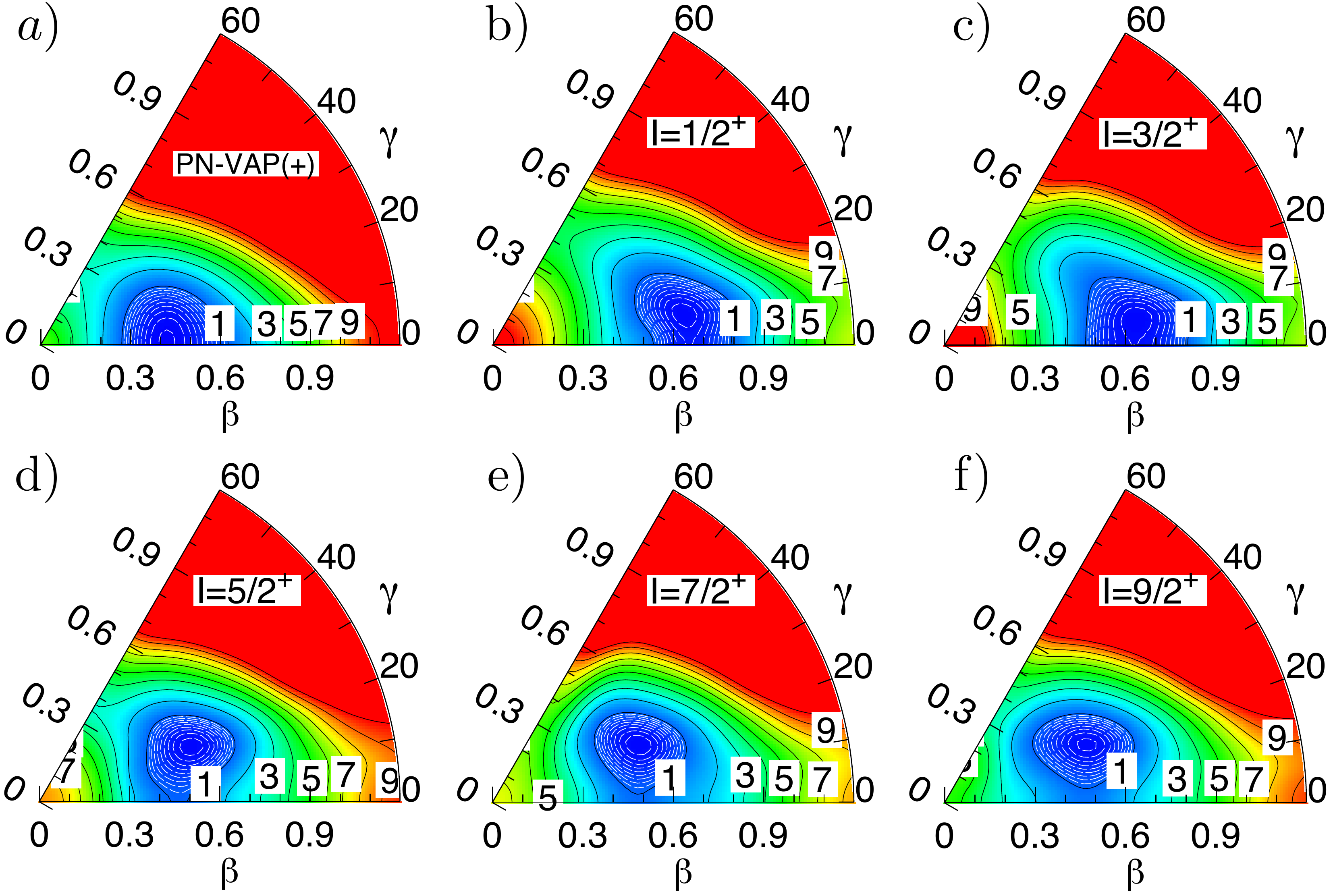}
 		\caption{Contour plots of the potential energy surfaces  as a function of $(\beta,\gamma)$ for positive parity. The panel a) stands for the PNVAP approach (no angular momentum projection), see Eq.~(\ref{Eq:PES_VAP}).  The panels b)-f)  correspond  to the PNAMP  approximation,  see Eq.~(\ref{Eq:PES_PNAMP}), for the angular momentum $I$ quoted  in the insets. The  solid black contour lines go from 1 to 10 MeV  in steps of 1 MeV. The dashed white lines start at zero and increase  by 0.1 MeV. The zero energy contour is only present if the minimum is flat enough. The angle $\gamma$ is given in degrees. In each panel the energies are relative to the corresponding energy minimum.}
 		\label{Fig:PES_POS}       
 	\end{center}
 \end{figure}
 
 \subsection{Potential energy surfaces}

One can obtain a great deal of information having a glance at the calculations at different stages of our procedure. The first piece of information is provided by the solution of the PNVAP equations, Eq.~(\ref{E_Lagr_bet-gam}), which determine the intrinsic wave functions $\tilde{\phi}^{\pi}(\beta,\gamma)$
for positive and negative parity.  	The PNVAP  equations have been solved in the $(\beta,\gamma)$ grid  mentioned above. The associated energies are given by
 \begin{eqnarray}   
{E}^{N,\pi}(\beta,\gamma)= \frac{ \langle\tilde{\phi}^{\pi}(\beta,\gamma)|\hat{H}\hat{P}^{N}|\tilde{\phi}^{\pi}(\beta,\gamma) \rangle}{\langle\tilde{\phi}^{\pi}(\beta,\gamma)|\hat{P}^{N}|\tilde{\phi}^{\pi} (\beta,\gamma)\rangle}. \label{Eq:PES_VAP} \end{eqnarray}
For the positive parity case  these energies are plotted in  panel $a)$ of Fig.~\ref{Fig:PES_POS} as contour plot in the $(\beta,\gamma)$ plane.  We  observe a well defined axially symmetric nucleus $(\beta\approx 0.42)$ which is rather soft in the $\gamma$ degree of freedom.  This softness is in agreement with the downsloping  character of the $1{\rm d}_{5/2}$ levels seen in the oblate part of  Fig.~\ref{Fig:SPE}.  The states that conform this potential energy surface (PES) do have good parity and particle number but are not eigenstates of the angular momentum.  Starting with the wave functions 
$\hat{P}^{N}|\tilde{\phi}^{\pi} (\beta,\gamma)\rangle$ one can obtain eigenvalues of $\vec{I}$  by
\begin{equation}
|\Phi^{N,I,\pi}_{M,\sigma } (\beta,\gamma) \rangle =   \sum_{K} F^{I}_{K\sigma} P^N P^I_{MK}|\tilde{\phi}^{\pi} (\beta,\gamma)\rangle, \label{Eq:Red_HW}
\end{equation}
with the variational  coefficients $F^{I}_{K\sigma}$  determined by the solution of a reduced Hill-Wheeler-Griffin equation,  obtained from  Eq.~(\ref{HW_Eq}) just by omitting  $\beta$ and $\gamma$ as running indices. This equation must be solved  at each point of the grid and for each angular momentum, see Ref.~\cite{BE.17} for further details. The PN and AM projected PES is given by
 \begin{eqnarray}   
{E}^{N.I\pi}_{\sigma}(\beta,\gamma)= \frac{ \langle\Phi^{N,I,\pi}_{M,\sigma } (\beta,\gamma) |\hat{H}|\Phi^{N,I,\pi}_{M,\sigma } (\beta,\gamma)  \rangle}{\langle\Phi^{N,I,\pi}_{M,\sigma } (\beta,\gamma) |\Phi^{N,I,\pi}_{M,\sigma } (\beta,\gamma) \rangle}. 
\label{Eq:PES_PNAMP} 
\end{eqnarray}

\begin{figure}[t]
	\begin{center}
		\includegraphics[angle=0,scale=.27]{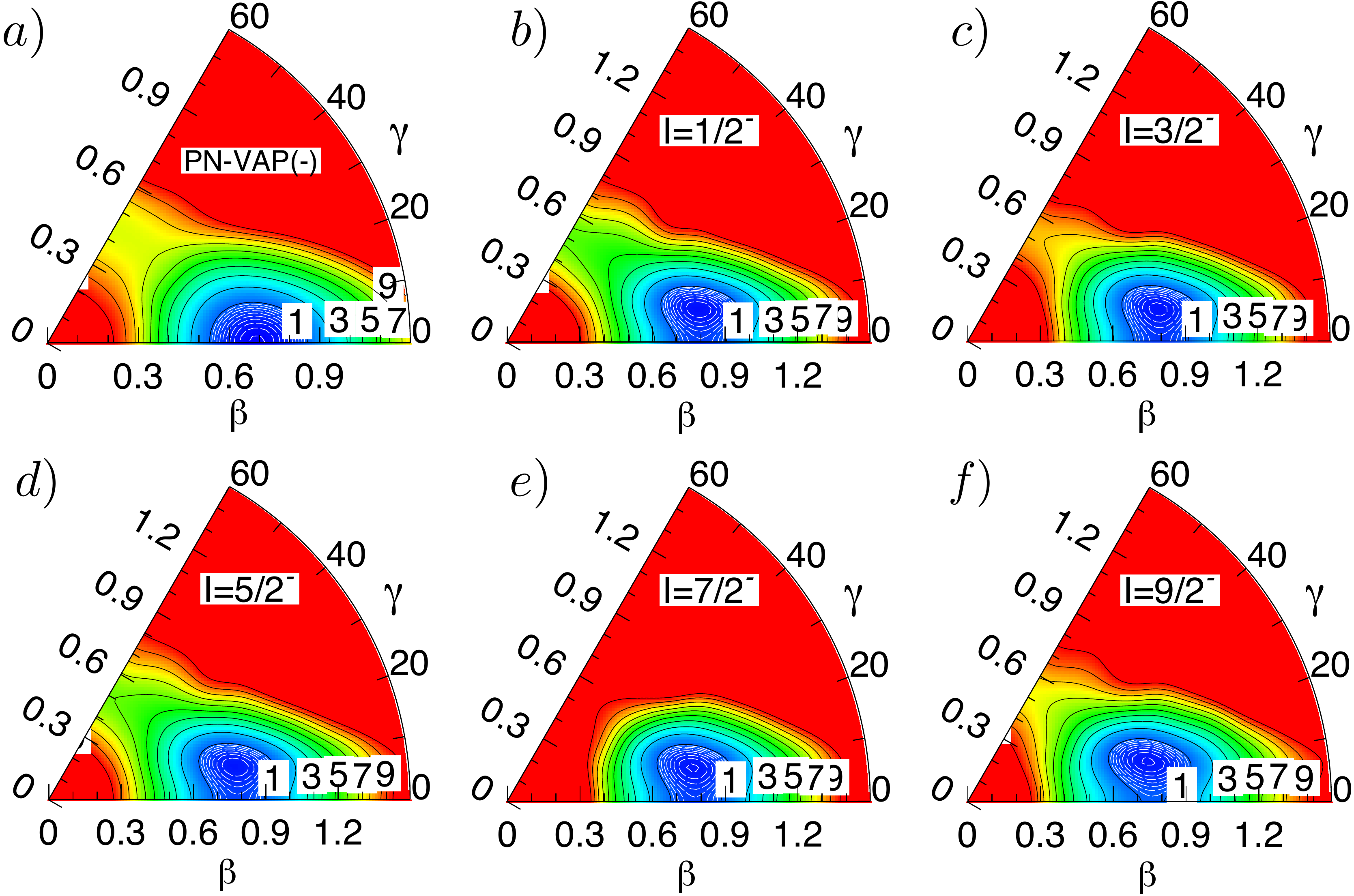}
		\caption{The same as Fig.~(\ref{Fig:PES_POS}) but for negative parity.}
		\label{Fig:PES_NEG}       
	\end{center}
\end{figure}

For a given value of $I$  the lowest energy corresponds to $\sigma~=1$.  The corresponding energies are plotted in  
 panels $b)-f)$ of Fig.~(\ref{Fig:PES_POS}).    At this stage the absolute minimum corresponds to  $I=5/2^{+}$ and  the relative energies of the other minima are $0.637$, $1.023$ and $1.522$ MeV for $I=1/2^{+}, 3/2^{+}$ and $7/2^{+}$, respectively.  In these plots we find that the angular momentum conservation shifts the minima to larger deformations. This is a well known effect of the angular momentum projection observed long ago \cite{RER.04,RE.07}. 
\begin{figure}[b]
	\begin{center}
		\includegraphics[angle=0,scale=.25]{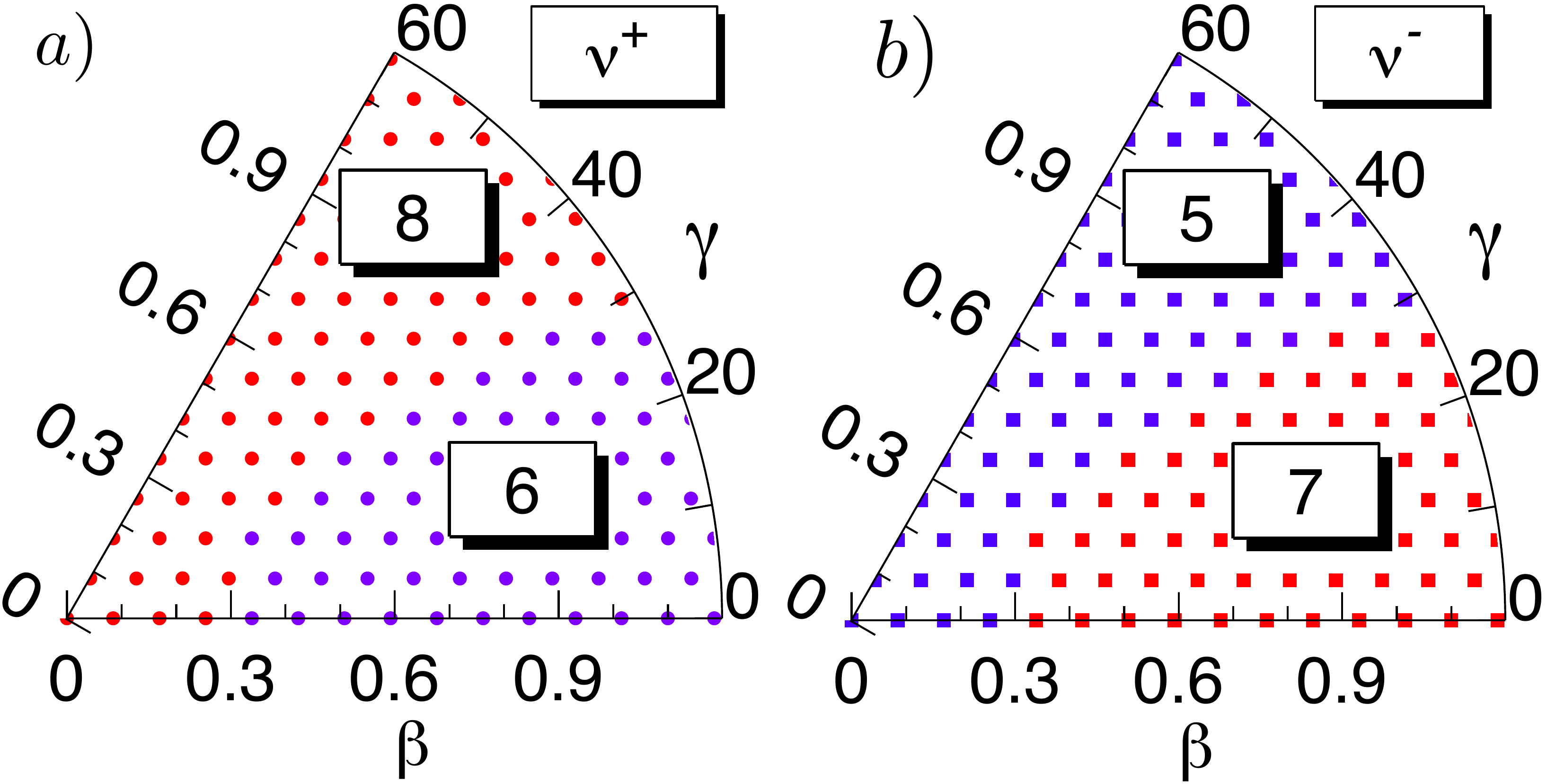}
		\caption{The expectation value of the particle number operator calculated with the intrinsic wave function of Eq.~(\ref{eq:odd_even_ansatz}) for positive-  (negative-) parity neutrons panel a) (panel b)) at each point of the $(\beta,\gamma)$ plane. See the text for details.}
		\label{Fig:NeutPlusMinus}       
	\end{center}
\end{figure}

\begin{figure*}[t]
\begin{center}
	\includegraphics[angle=0,scale=.87]{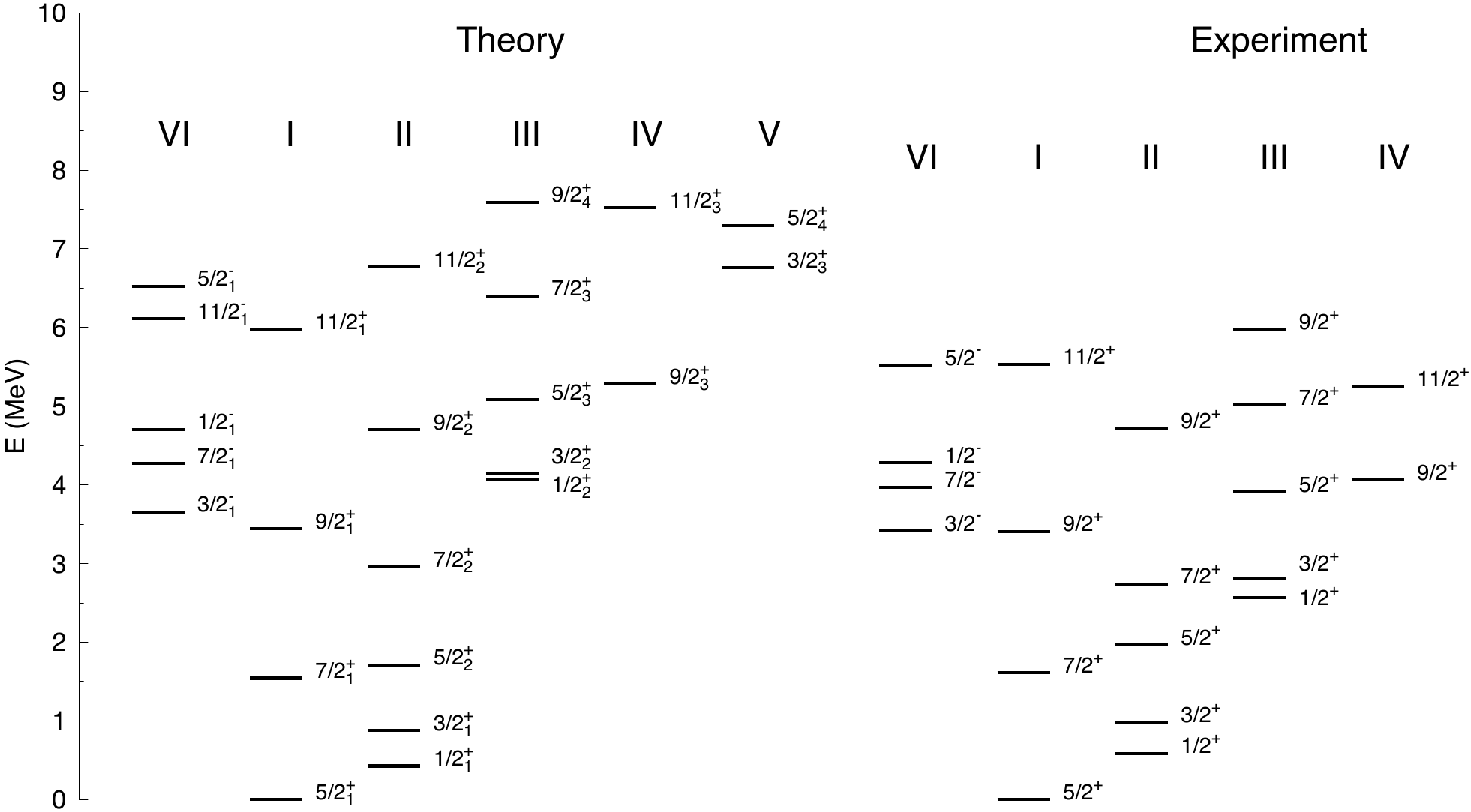}
	\caption{The spectrum of $^{25}$Mg from theory  (left) and  experiment(right), Ref.~\cite{Fi.09}.}
	\label{Fig:espectra}       
\end{center}
\end{figure*}

 As a function of $I$  we observe  in Fig.~\ref{Fig:PES_POS} two different regimes of the energy minima:  for  $I\ge 5/2\; \hbar $ we find smaller deformations and larger triaxialities  $(\beta\approx 0.51,\gamma \approx 25^\circ)$ than for  $I=1/2\;\hbar$  $(\beta\approx 0.55,\gamma \approx 7.6^\circ)$  and $3/2\;\hbar$ $(\beta\approx 0.64,\gamma \approx 6.6^\circ)$. We can get some insight into this comportment looking  at the  $F^{I}_{K\sigma}$ coefficients of  Eq.~(\ref{Eq:Red_HW}) at the energy minimum of each panel.  For   $I\ge 5/2\; \hbar $ we find that all these states are $K=5/2$ to a high degree of purity (at least $98\%$). In contrast,  for $I=1/2\;\hbar, 3/2\;\hbar$ they are pure  $K=1/2$. The fact that in the $sd$ shells, and in particular in the magnesium isotopes,  $K$ is practically a good quantum number  even if $\gamma \ne 0^\circ$  has been observed in earlier publications \cite{BE.17,RE.10}. This seems 
 to be a consequence  of the low single-particle level density which prevents a larger $K$-mixing. Notice that in a low-level-density regime we are in the strong coupling limit of the particle-plus-rotor model, where $K=\Omega$ is a good quantum number.
 A glance at  Fig.~\ref{Fig:SPE}  reveals than for $\beta\approx 0.52$ a level crossing between the levels [202 5/2] and  [211 1/2] takes place.  If $K$ is a good quantum number an odd neutron in the level  [202 5/2]  must have  $I\ge 5/2\; \hbar $.  After the crossing the odd neutron sits in the orbital [211 1/2] and can have any $I$ value. These $K=1/2$ states are higher in energy. The smaller $\gamma$ values of the $K=1/2$ states are a consequence of the fact that large deformations inhibit strong triaxialities.

In  Fig.~(\ref{Fig:PES_NEG}) we display the PES corresponding to the blocking of a neutron  orbital of negative parity. In panel a) we show the PNVAP results.  The energy minimum is axially symmetric and compared to  its positive parity counterpart  lies 4.237 MeV  higher in energy  and  has  a larger deformation,  $\beta=0.68$. Both results are to be expected if one considers the spe levels of Fig.~\ref{Fig:SPE}.   
In this channel there is a competition of   hole and particle states to host the odd neutron. These are the 
[101 1/2] orbital ($1{\rm p}_{1/2}$ sub-shell) and the [330 1/2] ([303 7/2]) in the prolate (oblate) branch  ($1{\rm f}_{7/2}$) sub-shell, respectively.  The dichotomy particle-hole allows a simple way to identify if
a particle or a hole is preferred in the variational process. 
The intrinsic wave function $|\tilde{\phi}^{\pi}\rangle$ of   Eq.~(\ref{eq:odd_even_ansatz}), solution of Eq.~(\ref{E_Lagr_bet-gam}),  factorizes in the form $|\tilde{\phi}_{p^+}\rangle |\tilde{\phi}_{p^-}\rangle |\tilde{\phi}_{n^+}\rangle |\tilde{\phi}_{n^-}\rangle $. The expectation values $\langle \tilde{\phi}_{n^+}|\hat{N}| \tilde{\phi}_{n^+} \rangle$ and $\langle \tilde{\phi}_{n^-}|\hat{N}| \tilde{\phi}_{n^-} \rangle$ provide us the number of neutrons with positive and negative parity.  These quantities are plotted in panels a) and b) of Fig.~\ref{Fig:NeutPlusMinus}, respectively.  In both panels we observe two well differentiated regions.  The one is a band along the oblate axis including spherical shapes represented by red (light gray) symbols in panel a) and blue (dark gray) symbols in panel b).  The other region corresponds to the rest of the $(\beta,\gamma)$ plane. In the upper part the number of neutrons with positive (negative) parity is 8 (5), i.e., the  upper areas correspond to configurations with a neutron hole in the [101 1/2] orbital. The lower parts with 6 (7) neutrons with positive (negative) parity correspond to configurations with a neutron particle  in the [330 1/2] orbital.  In the oblate area of  panel  b) we do not have states with 7 particles. We therefore conclude that the orbital [303 7/2] is never populated in the lowest configurations, i.e., it is more favorable to make a hole in the   [101 1/2] orbital. In the well prolate area, however, it is easier to put a particle in the [330 1/2] orbital.  Since the blocking structure is not affected by the angular momentum projection, this  picture provides a simple way to identify the components of the different wave functions. 

Resuming the discussion of panel a) of Fig.~\ref{Fig:PES_NEG} we find that the large $\beta$ deformation found in panel a) inhibits considerably triaxial  softness. We nevertheless observe an opening of the contour  lines at $\beta \approx 0.5$ on the oblate side.  These shapes correspond to configurations with a hole in the [101 1/2],  see Fig.~\ref{Fig:SPE} and Fig.~\ref{Fig:NeutPlusMinus}.  The orbital [101 1/2] is preferred as compared to the  [303 7/2]   because the former is upsloping and 
gets closer to the Fermi level with increasing deformation. 
In panels b)-d) of the same figure we depict the angular momentum PES of Eq.~(\ref{Eq:PES_PNAMP}).  As compared with the PNVAP results here we observe, as in the positive parity case, a shift to larger deformations and to triaxial shapes.  In this case, however, the shift towards triaxial shapes is smaller owing to the fact that the $\beta$ deformations are larger than in the  positive parity case.  The energy minimum remains in the same position, around  $\beta=0.77,\gamma=11^\circ$ for all $I$-values. This location corresponds, see  Fig.~\ref{Fig:NeutPlusMinus}, to a particle in the [330 1/2] orbital.  The $K$-composition of the minimum wave function, i.e., the   $F^{I}_{K\sigma}$ coefficients of  Eq.~(\ref{Eq:Red_HW}),  indicates that this assignment is correct   since they have a  very pure $K=1/2$, independently of their $I$-value.  
From Figs.~\ref{Fig:PES_NEG}, \ref{Fig:NeutPlusMinus} we conclude that the oblate configurations corresponding to holes in the [101 1/2] orbital ($1{\rm p}_{1/2}$ subshell)  are very high in energy.
The relative energies of the minima referred  to the $I=5/2^{+}$ energy are  $3.567$ MeV for the  $I=3/2^{-}$   followed by $4.228$  MeV, $4.605$ MeV, and $6.131$  for  $I=7/2^{-},1/2^{-}$ and $11/2^{-}$ states, respectively. 

We would like to remark that the plots of Figs.~\ref{Fig:PES_POS},\ref{Fig:PES_NEG} correspond to the $\sigma=1$ of the reduced HWG equation, higher lying (i.e. $\sigma \ge 2$) solutions may behave differently. Notice also that in the solution of the general HWG equation corresponding to the SCCM approach,  Eq.~(\ref{HW_Eq}), all $\sigma$ states are included.

\section{Symmetry Conserving Configuration Mixing Results} \label{Sect:SCCM}
 Once the basis states of the GCM wave function,  Eq.~(\ref{GCM_BASIS}), are determined the next step is the solution of Eq.~(\ref{HW_Eq}).  The mixing of the basis states includes the dynamical correlations providing   non-collective 
 and collective states  such as  the $\beta$ and $\gamma$  vibrations.

 For the   dynamical properties  pairing  correlations play a crucial role.  As a matter of fact it has been shown in Ref.~\cite{BE.18} that the  PN-VAP treatment in the solution of Eq.~(\ref{E_Lagr_bet-gam}) is relevant to obtain the super-fluid wave functions. In the  HFB plus PN projected approach, on the other hand,   the pairing collapse takes place  in many points of the $(\beta,\gamma)$ grid. This collapse happens  in weak pairing situations which is very often encountered in odd-A nuclei due to the blocking effect.

  The SCCM calculations are rather lengthy because the configuration mixing implies the calculation of $N(N+1)/2$  matrix overlaps, with $N$ the number of grid points  in the $(\beta,\gamma)$ plane. To alleviate these calculations it is necessary to restrict as much as possible the number of points. Taking into account that the energies of the points forming the red color areas of the PES's of Figs.~\ref{Fig:PES_POS},\ref{Fig:PES_NEG} are energetically very high with respect to the energy minimum,  one can expect that they will not mix very much with the lower lying-ones.   Therefore
 for the positive-parity case we restrict the calculations to 81 wave functions in the range $0\le \beta \le 1.1$, $0\le \beta \sin\gamma \le 0.45$.  In the negative parity calculations we extend the maximal $\beta$-value up to 1.4 which gives 95 grid points.

 The solution of Eq.~(\ref{HW_Eq}) provides the eigenvalues $E^{N,I,\pi}_{\sigma}$ and eigenfunctions $|\Psi^{N,I,\pi}_{M,\sigma }  \rangle$.  Properties like transitions, quadrupole moments and so on,  together with the collective wave functions, Eq.~(\ref{coll_wf}),  allow to build up  the excitation spectrum as well as the interpretation of the different states.  
We can clearly identify 5 bands of positive parity, namely the ground band or $5/2^{+}_1$ band (which we will call band I),  the first excited 1/2$^+$ or   $1/2^{+}_1$ band (band II), the second excited 1/2$^+$ or  $1/2^{+}_2$ band, (band III), the 9/2$^+$  band (band IV)  and the 3/2$^+$ band (band V). Additionally we  identify a  
negative parity band (band VI), with a $I=3/2^{-}$ state as band head. 
  These bands are displayed on the left hand side of Fig.~\ref{Fig:espectra} together with the corresponding experimental ones (except band V)  on the right hand side. 

  Before discussing these results we would like to comment on the weak and the strong points of our approach. We have mentioned in  Subsection.~\ref{susect:bl_eq} that by considering the wave functions of the  $(\beta,\gamma)$ plane a good approximation to a AM-VAP was reached with respect to these variables for which the AM projected energy shows a strong dependence. There is, however, a third variable which also shows an strong dependence, namely the alignment of pairs 
  	(or the cranking frequency) which has not been considered.
The  lack of an alignment dependence in the variational equations used to determine the HFB w.f.  favors  states with low angular momentum $I$ disfavoring thereby the higher  ones (the larger $I$ the more). The result is a stretched spectrum as compared with the experiment. Possible remedies to this situation are the consideration of the angular frequency as an additional generator coordinate as done in Refs.~\cite{BRE.15,EBR.16}
or the inclusion of additional  one-quasi-particle states, Eq.~(\ref{eq:odd_even_ansatz}) at each $(\beta,\gamma)$ point of the grid in the SCCM  ansatz of  Eq.~(\ref{GCM_ANSA}), see Ref.~\cite{CE.16,CE.17}. Both procedures allow to include aligned configurations at each $(\beta,\gamma)$ point.
On the other hand, our collective wave function, Eq.~(\ref{GCM_ANSA}), allows the mixing of different configurations, contrary to the Nilsson or Particle-plus-Rotor type calculations that necessarily assign a given orbital to each band.  As compared with shell model calculations, our ansatz allows to identify very clearly collective bands like $\beta$ or $\gamma$ bands.

\begin{figure}[t]
	\begin{center}
		\includegraphics[angle=0,scale=.26]{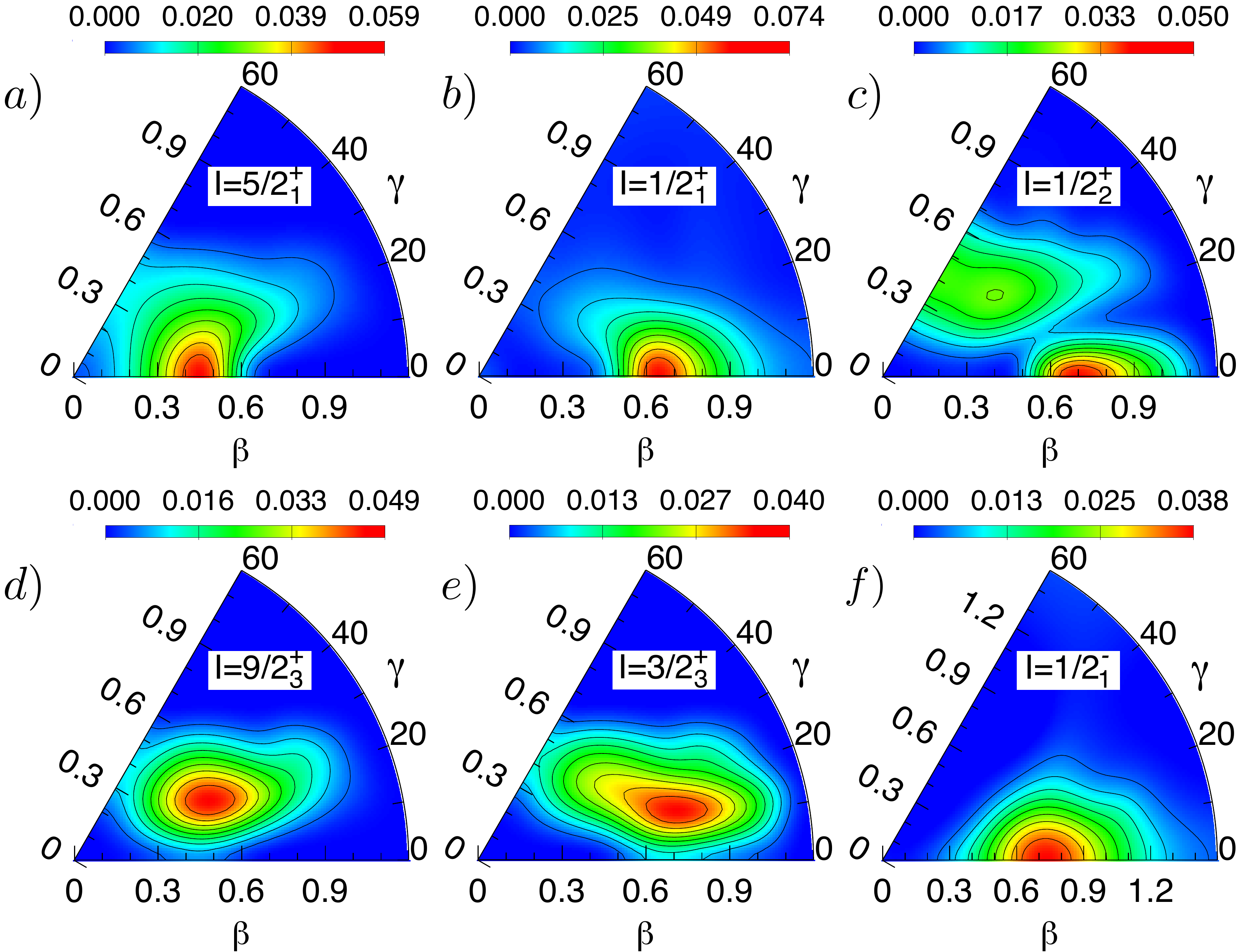}
		\caption{Squared collective wave functions of the band heads of $^{25}$Mg in the $(\beta,\gamma)$ plane.  The spin and parity of the
			different states is given in the inset of each plot.  In each plot  the value of the outer contour corresponds to one tenth of the maximum value shown in the corresponding palette.  Each contour
			is incremented by this amount up to the maximum value.  The angle $\gamma$ is given in degrees.}
		\label{Fig:WFS}       
	\end{center}
\end{figure}

    \begin{table} [b]
	
	\begin{tabular}{ccccccc}
		\hline
		$I_{\sigma}^{\pi}$    &\hspace{.05cm}$K=\pm1/2 $ &  $\hspace{.2cm}\pm3/2 $ &$\hspace{.2cm}\pm5/2 $ & $\hspace{.2cm}\pm7/2$ & $\hspace{.2cm}\pm9/2$ & $\hspace{.2cm}\pm11/2$
		\tabularnewline
		\hline
		$5/2^{+}_{1}$ & 1.6 & 1.4  &  97.0 & - &  - & -
		\tabularnewline
		$7/2^{+}_{1}$ & 0.6 & 0.2  &  98.8 & 0.4 &  - & -  
		\tabularnewline
		$9/2^{+}_{1}$ & 1.0 & 0.5  &  96.7 & 0.3 &  1.5 & - 
		\tabularnewline
		$11/2^{+}_{1}$ & 3.6 & 0.3  &  94.8 & 0.4 &  0.6 & 0.3
		\tabularnewline    		    		
		\hline
		$1/2^{+}_{1}$ & 100 & -  &  - & - &  - & -
		\tabularnewline
		$3/2^{+}_{1}$ & 99.6 & 0.4  &  - & - &  - & -  
		\tabularnewline
		$5/2^{+}_{2}$ & 99.4 & 0.1  &  0.4 & - &  - & - 
		\tabularnewline
		$7/2^{+}_{2}$ & 98.8 & 0.1  &  0.9 & 0.2 &  - & -
		\tabularnewline 
		$9/2^{+}_{2}$ & 96.5 & 0.6  &  2.0 & 0.2 &  0.8 & -
		\tabularnewline
		$11/2^{+}_{2}$ & 95.5 & 0.3  &  3.2 & 0.1 &  0.6 & 0.3
		\tabularnewline
		\hline
		$1/2^{+}_{2}$ & 100 & -  &  - & - &  - & -
		\tabularnewline
		$3/2^{+}_{2}$ & 96.5 & 3.5  &  - & - &  - & -  
		\tabularnewline
		$5/2^{+}_{3}$ & 91.3 & 6.1  &  2.6 & - &  - & - 
		\tabularnewline
		$7/2^{+}_{3}$ & 94.6 & 1.9  &  2.9 & 0.6 &  - & -
		\tabularnewline 
		$9/2^{+}_{4}$ & 80.2 & 8.6  &  9.8 & 0.7 &  0.7 & -
		\tabularnewline
		$11/2^{+}_{4}$ & 87.6 & 3.2  &  8.0 & 0.3 &  0.5 & 0.3
		\tabularnewline
		\hline   		
		$9/2^{+}_{3}$ & 0.8 & 0.3  &  1.0 & 0.2 &  97.6 & -
		\tabularnewline
		$11/2^{+}_{3}$ & 0.7 & 0.1  &  1.5 & 0.1 &  97.6 & 0.1
		\tabularnewline
		\hline		
		$3/2^{+}_{3}$ & 2.2 & 97.8  &  - & - &  - & -
		\tabularnewline
		$5/2^{+}_{4}$ & 2.9  & 93.1  &  4.0 & - &  - & -
		\tabularnewline
		\hline
		$1/2^{-}_{1}$ & 100 & -  &  - & - &  - & -
		\tabularnewline
		$3/2^{-}_{1}$ & 98.9 & 1.1  &  - & - &  - & -  
		\tabularnewline
		$5/2^{-}_{1}$ & 93.2 & 6.5  &  0.2 & - &  - & - 
		\tabularnewline
		$7/2^{-}_{1}$ & 96.5 & 3.5  &  0.0 & 0.0 &  - & -
		\tabularnewline 
		$11/2^{-}_{1}$ & 94.6 & 5.3  &  0.1 & 0.0 &  0.0 & 0.0
		\tabularnewline
		\hline
	\end{tabular}
	\caption{K-distribution of the different states}
	\label{tab:pesoK}
\end{table}

Looking at Fig.~\ref{Fig:espectra}  we find an overall good qualitative agreement between the two sets of bands. The similitude between both spectra  is specially good for bands I, II and VI. As we shall see below these bands correspond to different configurations. The theoretical results for bands III and V,  though providing the right level ordering, lie higher than the experimental counterparts. In general  our spectrum is a bit stretched as compared with the experimental one but, as shown in Refs.~\cite{BRE.15,EBR.16} for even-even nuclei, that can be corrected  if one considers the cranking frequency as an additional coordinate in the SCCM calculations. In particular, bands III, IV and V  are of vibrational character which in terms of the QRPA involves the consideration of additional quasi-particle excitations while in our calculation  we only consider the lowest blocked state at each $(\beta,\gamma)$ point.  This does not mean that we do not include these states.  We do it at other $(\beta,\gamma)$ values but they are higher in energy. The explicit consideration of these states lowers considerably the energy of the collective vibrations,   see Ref.~\cite{CE.16,CE.17}. Theoretical descriptions of bands in triaxial nuclei in the framework of the collective model can be found in Refs.~\cite{Da.61,MFK.97}.

 \begin{figure}[t]
	\begin{center}
		\includegraphics[angle=0,scale=.31]{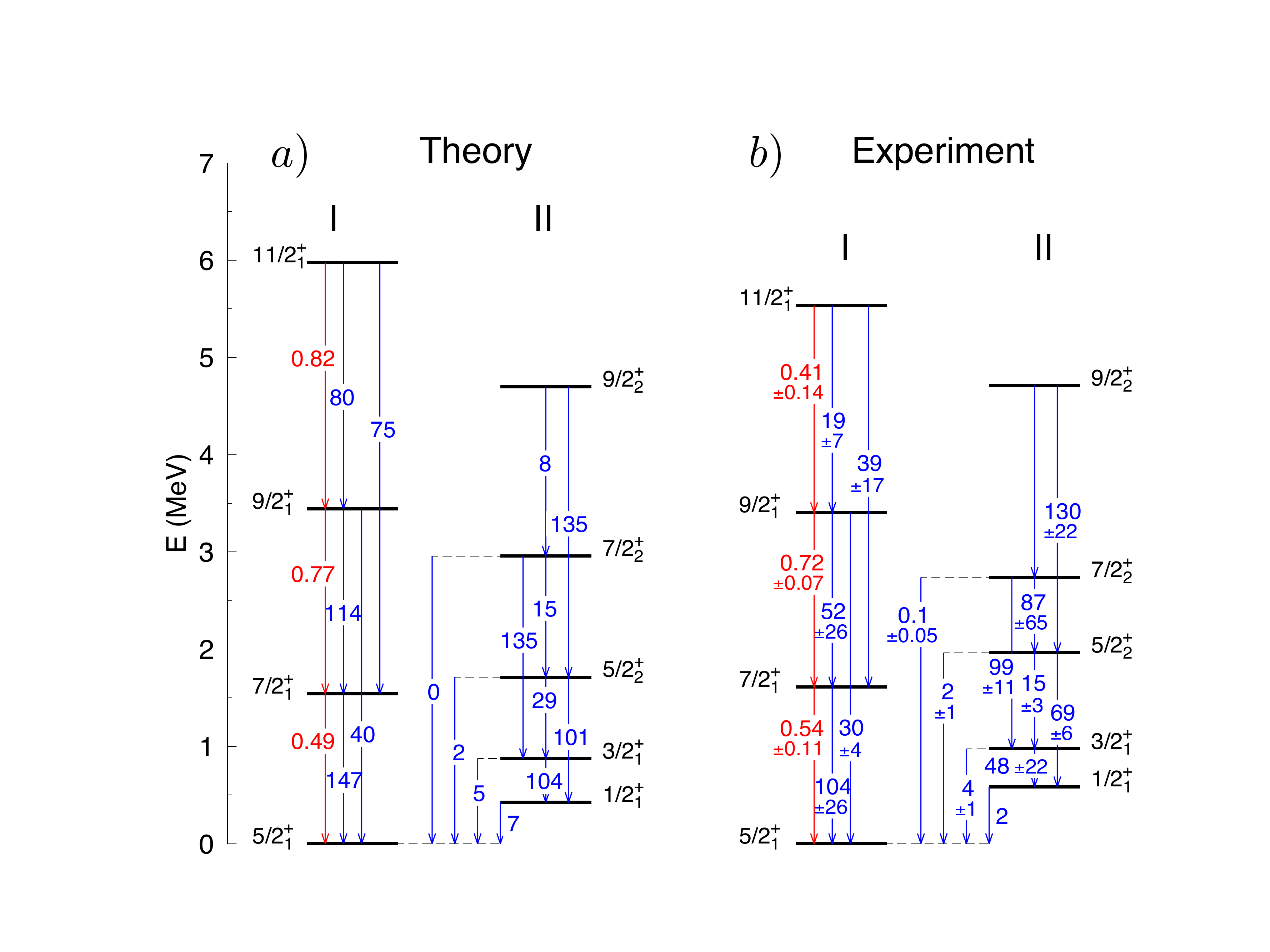}
		\caption{ Transition probabilities of the two lowest bands. In panel b)  the  experimental values \cite{Fi.09}
			are shown and in panel a) the  theoretical ones.  The numbers in blue (dark gray) color correspond to B(E2) values, in e$^2$fm$^4$,  and those in red (light gray)  to B(M1) ones, in  $\mu_N^2$.}
		\label{Fig:trans_1exc_gs}       
	\end{center}
\end{figure}

   In Fig.~\ref{Fig:WFS} we display the collective wave functions $|p^{\sigma I}_{K}(\beta,\gamma)|^2$ of Eq.~(\ref{coll_wf})  for  the band heads of the spectrum shown in Fig.~\ref{Fig:espectra}. The  wave functions of the excited states of each band, not shown here, do not differ much from their corresponding band heads.

   \begin{table}[t]
	\centering
	
	\begin{tabular}{cccccc}
		\hline
		$I_{\sigma}^{\pi}$ &\hspace{.7cm}$\overline{\beta}^{I\pi\sigma}$ & \hspace{.7cm} $\overline{\gamma}^{I\pi\sigma}$  & \hspace{.25cm}$Q_{\rm spec.} $& \hspace{.25cm}$Q_{\rm spec.}^{SM}$& \hspace{.25cm}$Q_{\rm spec.}^{RM}$
		\tabularnewline
		\hline
		$5/2^{+}_{1}$ & 0.505 & 21.16  & 22.2 & 20 &20
		\tabularnewline
		$7/2^{+}_{1}$ & 0.523 & 18.96 & 3.8& 3&3-7
		\tabularnewline
		$9/2^{+}_{1}$ & 0.515 & 18.94  & -6.8& 9& -5
		\tabularnewline
		$11/2^{+}_{1}$ & 0.542 & 18.73  & -11.6& -&-
		\tabularnewline    		    		
		\hline
		$1/2^{+}_{1}$ & 0.669 & 12.87 & 0.0& 0&0
		\tabularnewline
		$3/2^{+}_{1}$ & 0.687 & 12.42   & -14.8&-13 &-11
		\tabularnewline
		$5/2^{+}_{2}$ & 0.674 & 11.92  & -20.2&-15 &-16
		\tabularnewline
		$7/2^{+}_{2}$ & 0.708 & 9.39 & -25.3&-21 &-18
		\tabularnewline 
		$9/2^{+}_{2}$ & 0.666 & 11.75 & -25.2&-17 &-20
		\tabularnewline
		$11/2^{+}_{2}$ & 0.728 & 8.28  & -30.3& -&-
		\tabularnewline
		\hline
		$1/2^{+}_{2}$ & 0.639 & 23.66  & 0.0& -&-
		\tabularnewline
		$3/2^{+}_{2}$ & 0.638 & 22.72  & -14.0&-11 &-11
		\tabularnewline
		$5/2^{+}_{3}$ & 0.665 & 20.39  & -16.8& -15&-16
		\tabularnewline
		$7/2^{+}_{3}$ & 0.630 & 23.27  &-23.0&-16 &-18
		\tabularnewline 
		$9/2^{+}_{4}$ & 0.659 & 19.26  & -18.6& -&-
		\tabularnewline
		$11/2^{+}_{4}$ & 0.630 & 20.83  & -26.2& -&-
		\tabularnewline
		\hline    		
		$9/2^{+}_{3}$ & 0.592 & 26.72  & 35.5& 18&30
		\tabularnewline
		$11/2^{+}_{3}$ & 0.604 & 24.88  & 15.7& -&-
		\tabularnewline
		\hline  		
		$3/2^{+}_{3}$ & 0.699 & 22.65  & 14.8& &
		\tabularnewline
		$5/2^{+}_{4}$ & 0.684  & 23.01 & -5.3& &
		\tabularnewline
		\hline
	\end{tabular}
	\caption{Average deformation parameters $\overline{\beta}^{I\pi\sigma}$  and $\overline{\gamma}^{I\pi\sigma}$ together with the spectroscopic quadrupole moment in ${\rm e\; fm^2}$ in columns 4, 5 and 6 ( this work, shell-model, SM, and rotational model, RM, respectively)  for the different states. The SM and RM values are taken from Ref.~\cite{Cole.75}.}
	\label{tab:promedio}
\end{table}
  
  For an interpretation of these bands we present in Table~\ref{tab:pesoK} the $K$ distributions of the different states calculated according to Eq.~(\ref{p1}).  Interestingly there is little mixing and the quantum number $K$ is rather pure. As mentioned before this is due to the low single particle level density in light nuclei and to the large $\beta$ deformation of this nucleus, i.e., we are in the strong coupling limit.

  \subsection{The ground band  (band I)}
  The ground state of $^{25}$Mg has  $I=5/2^{+}$ and the members of the  ground band,  band I in Fig.~\ref{Fig:espectra}, are nearly pure $K=5/2$  ($\ge 95\%$), see Table~\ref{tab:pesoK}.
In Table~\ref{tab:promedio} we show the average $\beta$ and $\gamma$ values, calculated with the help of Eq.~(\ref{p2}),  together with the spectroscopic quadrupole moments. The  $\overline{\beta}$ value of the ground state is $0.505$.  A look at Fig.~\ref{Fig:SPE} indicates that this rotational band is based on the [202 5/2] orbital. This assignment is consistent with the ones found in the literature \cite{Headly.88}. The collective wave function of Eq.~(\ref{p2}) is represented in panel a) of Fig.~\ref{Fig:WFS}. If we compare this plot with the corresponding PES, i.e., panel d) of Fig.~\ref{Fig:PES_POS}, we can observe the dynamical effects introduced by the configuration mixing. Thus, though the maximum of the distribution is  approximately at $\beta=0.45$, i.e, similar to the energy minimum of the PES, it is shifted to axially symmetric shapes. For $\beta > 0.6$ we observe a sharp decrease of the probability values.  This is a clear indication that the Hamiltonian matrix elements of the states
$P^N P^I_{MK} |{\tilde\phi}^{\pi} (\beta,\gamma)\rangle$ of Eq.~(\ref{GCM_BASIS})  based on the orbitals 
 [211 1/2] and  [202 5/2],  which cross at  $\beta\approx 0.5$ (see Fig.~(\ref{Fig:SPE})),  are less attractive than others inhibiting the mixing of the  states based on the  [211 1/2] configuration in the collective wave function. 
    
    In Fig.~\ref{Fig:trans_1exc_gs} we display the theoretical and experimental reduced transition probabilities B(M1) and B(E2) along the ground band (band I).  In general there is a good overall agreement between theory and experiment. The theoretical B(E2) values are somewhat larger than the experimental ones as it  is also the case for even-even nuclei \cite{RE.10}, while the agreement for the magnetic transitions is much better. 
    
 The spectroscopic quadrupole moments of the ground band listed in Table~\ref{tab:promedio} are in good agreement with both  the shell model  and the rotational model \cite{Cole.75}.

 \subsection{The first excited $1/2^{+}$ band (band II)} 
 
   The first excited band (band II in Fig.~\ref{Fig:espectra}) is based on a state with $I=1/2^+$.  It is again a very pure band, see Table~\ref{tab:pesoK}, and its average $\beta$ value is 0.67, see Table~\ref{tab:promedio}.  For this value we find from Fig.~\ref{Fig:SPE} that the only $\Omega=1/2$ orbital available  around this $\beta$ value is [211 1/2].  Band II  is a rotational band built on this orbital. It cannot be a collective excitation of the ground band because the transition probabilities connecting both bands are very small, as a matter of fact the 1/2$^+$ band head is an isomeric state ($T_{1/2}= 3.3\; {\rm ns})$  \cite{Fi.09}.  The corresponding collective wave function is represented in panel b) of Fig.~\ref{Fig:WFS}.  Its maximum is located in the minimum of the PES plot (see panel b) of Fig.~\ref{Fig:PES_POS}). The wave function is rather concentrated around its maximum indicating the non-collective character of the state.  In this plot, obviously, we do not observe   the drop in probability density for $\beta>0.6$ observed in the wave function of the ground state.

     In Fig.~\ref{Fig:trans_1exc_gs} we display the theoretical and experimental reduced transition probabilities  B(E2) along the band.  Since this band is a $K=1/2$ band the B(M1) transition probabilities are smaller than for the ground band. The larger deformation of this band provides a good rotational band which is  somewhat distorted by the decoupling parameter due to its   $K=1/2$ character. The theoretical B(E2) values are, again, somewhat larger than the experimental ones.
     In  Fig.~\ref{Fig:trans_1exc_gs} the decay from the 1/2$^+_1$ to the ground band is also shown.  The agreement between theory and experiment is  very good with the exception of the E2 transition $1/2^+_1\rightarrow  5/2^+_1$. The  experimental value is  2.44 e$^2$fm$^4$ and the theoretical one 7.3  e$^2$fm$^4$.  Notice that since  the ground band is  rather pure $K=5/2$ and the 1/2$^+_1$ band pure  $K=1/2$, there are no M1 transitions between the members of the two bands.
     
   The spectroscopic quadrupole moments of this band, see Table~\ref{tab:promedio},  are also in good agreement with the SM and the rotational RM \cite{Cole.75}.

   \subsection{The second excited $1/2^{+}$ band (band III)}

    The second  excited band (band III in Fig.~\ref{Fig:espectra}) has as  band head an $I=1/2^+$ state,  its average deformation is $\beta=0.639$. 
    It is also a very pure $K=1/2$ band. If this band were based on a pure particle-hole excitation, the next available $\Omega =1/2$ orbital would be the [200 1/2].  This assignment has been made by some authors \cite{BM.75,Endt.98} but there are also collective model studies, which assigned to this band a mixed character of the [200 1/2] state and a  $(K-2)_{\gamma}$ vibration on the [202 5/2] state \cite{Headly.88} .   As we can see in Fig.~\ref{Fig:espectra}, the theoretical values for the energies of this band are a bit high as compared with the experimental values.  This is probably due to the fact that with our blocking procedure we can only block  the orbital [200 1/2]  through the pairing correlations which provide a given probability to populate this orbital, see Fig.~\ref{Fig:SPE},  or  through mixing in the $(\beta,\gamma)$ plane.  
    
       \begin{figure}[h]
      	\begin{center}
      		\includegraphics[angle=0,scale=.27]{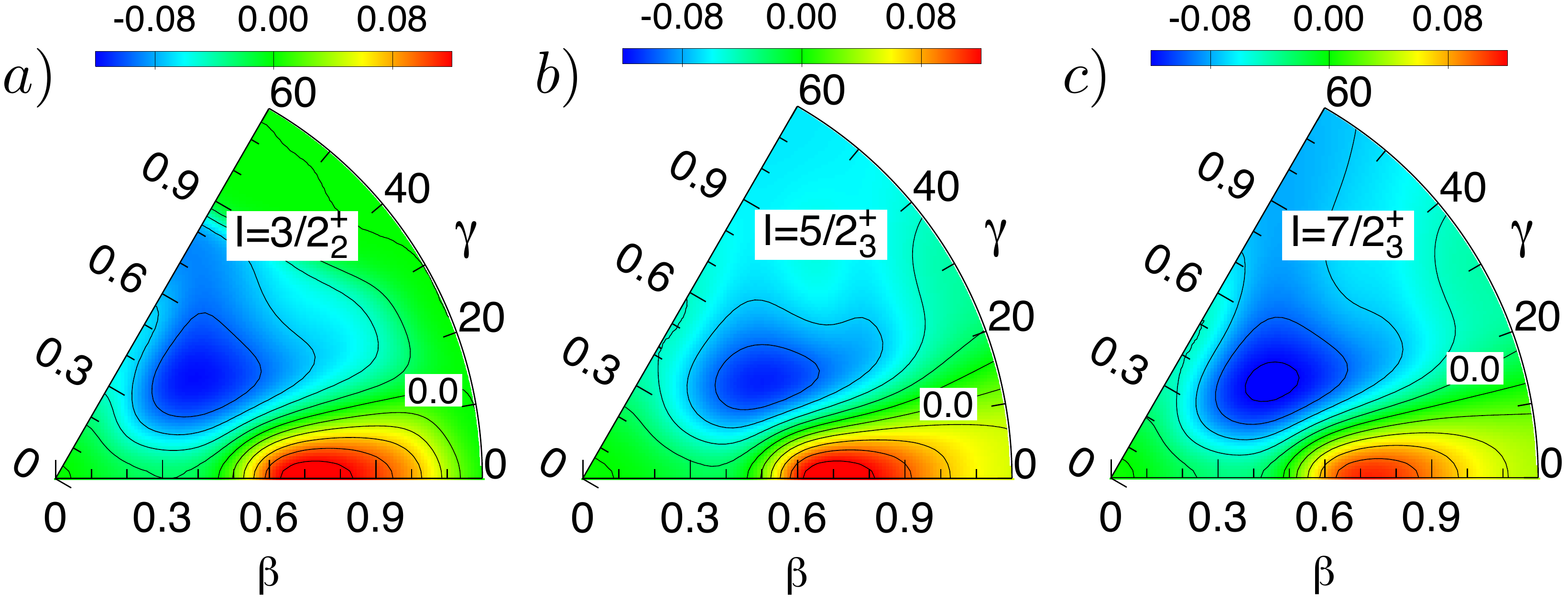}
      		\caption{Collective wave functions of the excited states of band  III  of $^{25}$Mg in the $(\beta,\gamma)$ plane with their signs.  The spin and parity of the
      			different states is given in the inset of each plot and will be used to label  the plot. The angle $\gamma$ is given in degrees.}
      		\label{Fig:gamma_nodal}       
      	\end{center}
      \end{figure}

      \begin{figure}[h]
      	\begin{center}
      		\includegraphics[angle=0,scale=.32]{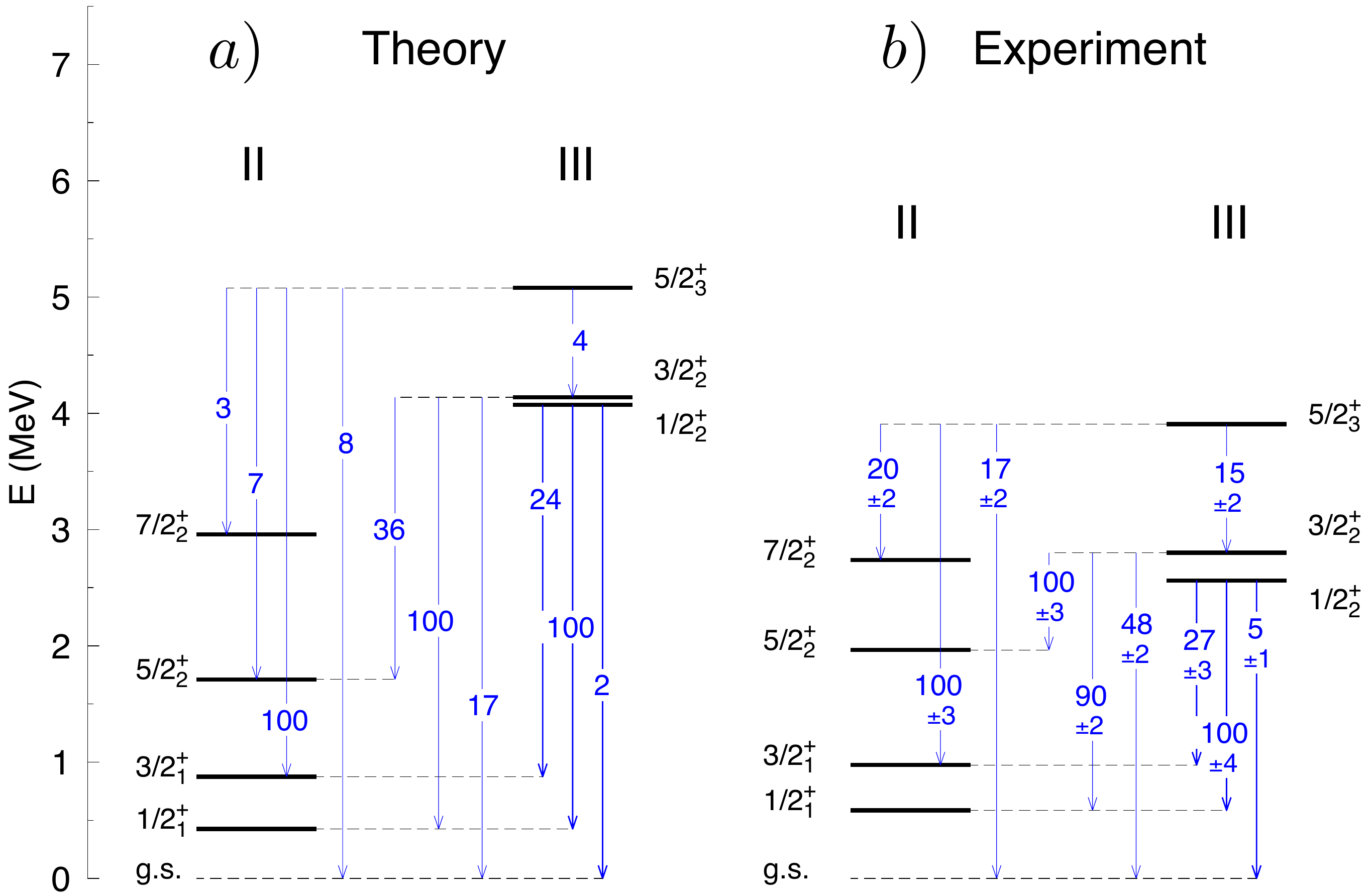}
      		\caption{Branching ratios of the $1/2_{2}$ band, theory (panel a)) and experiment (panel b)).}
      		\label{Fig:trans_11_12}       
      	\end{center}
      \end{figure}
      
       The wave function  of the band head is plotted in panel c) of Fig.~\ref{Fig:WFS}. It has two peaks, one at $\beta\approx 0.7, \gamma=0^\circ$ and the other at $\beta\approx 0.45, \gamma\approx 50^\circ$, the nodal line in-between corresponds to the $n_\gamma =1$ character of the vibration. The fact that the second peak appears at a smaller $\beta$ value is due to the energy rise of the [211 1/2] and  [200 1/2] orbitals on the oblate side for large $\beta$ values, see Fig.~\ref{Fig:SPE}. To illustrate more clearly the $n_{\gamma}=1$ character of this band we have plotted in Fig.~\ref{Fig:gamma_nodal} directly the collective  wave function  $p^{\sigma I}_{K}(\beta,\gamma)$ of Eq.~(\ref{coll_wf}), not its module as in Fig.~\ref{Fig:WFS}, for the  $I=3/2^{+},  5/2^{+}$ and $7/2^{+}$ members of this band, panels a), b) and c) respectively. The three wave functions are rather similar as expected for a rotational band. Here we clearly see the nodal line separating the positive and negative contours. Additional information on the nature of the band is provided by the branching ratios for the decay of its members. These branching ratios, normalized in each case to the strongest branch, are shown in  Fig.~\ref{Fig:trans_11_12}  where both the experimental, panel b), and theoretical values, panel a), are given.  The theoretical branching ratios are evaluated using the calculated reduced transition probabilities and the experimental $\gamma$-ray energies. The agreement between theory and experiment is good and as one can see the decay proceeds almost exclusively to the first excited  $1/2^+_1$ band  with only a  small strength to the ground state.  In particular, the $1/2^{+}_2$ level decays with a value of 100 to the band head of band II, 24 to the    $3/2^{+}_2$ level and 2 to the ground state, while the corresponding experimental values are 100, 27 and 2,  indicating a   very good  agreement.   Unfortunately  there are no separated experimental values for all magnetic and electric transitions. For the particular case of the $1/2^{+}_2 \rightarrow 5/2^{+}_1$ E2 transition   to the ground state, a value of  $19\pm 13$ e$^2$fm$^4$  has been measured which compares well with  our theoretical value of 13 e$^2$fm$^4$.  For the other members of the band the agreement with the experiment is not so good as for the band head but the main features are correctly described. The fact that the $1/2^+_2$ state decays mainly to the  $1/2^{+}_1$ level suggest the assignment of this band as a $\gamma$-band,   $K=0, n_{\gamma}=1$,  built on the  $1/2^{+}_1$ band. Notice that this assignment is only possible because this nucleus is triaxial \cite{Da.61,MFK.97}
       
        One could now ask about the mean field interpretation \cite{BM.75,Endt.98} of assuming the $1/2^+_2$  as a rotational band build on the  [200 1/2] orbital.  First, they are based on axially symmetric Nilsson calculations and second,  that in the sd shells there is a lot of mixing, and while  for the  $1/2^{+}_1$ band (band II) the occupation probability is
        large for  the orbital [211 1/2] and small for the  [200 1/2],  for the $1/2^+_2$ band (band III) it is the other way around.

      The spectroscopic quadrupole moments of this band are also in good agreement with the results of SM and the RM calculations see Table~\ref{tab:promedio}.

   \begin{figure}[h]
   	\begin{center}

   		\includegraphics[angle=0,scale=.31]{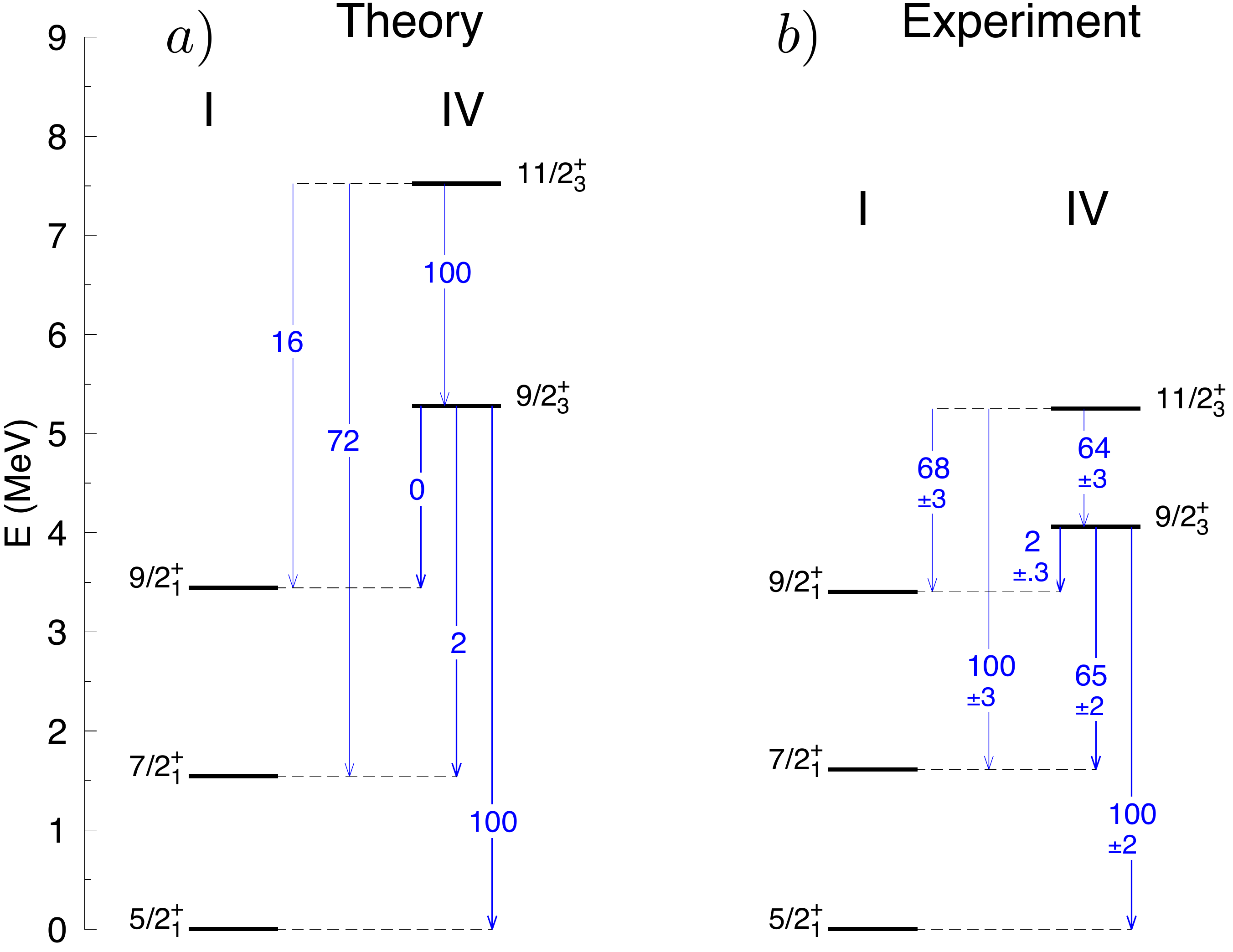}
   		\caption{Branching ratios for the decay of band IV,  in panel a)  the theory and in panel b)  the experimental data according to the
   			addopted values of Ref.~\cite{Fi.09}.}
   		\label{Fig:band_93}       
   	\end{center}
   \end{figure}

 \subsection{The 9/2$^+$ excited band (band IV)}
   The third excited band (band IV) has an  $I=9/2^+$ band head. It is a rather pure $K=9/2$ band with average deformation parameters $\beta= 0.592$ and $\gamma=26.7^{\circ}$.  Since there is no single-particle state of these characteristics around the Fermi level it is clear that it has to be a collective band.  In Ref.~\cite{BM.75}  it has been interpreted as a $(K+2)_{\gamma}$ vibration on  the [202 5/2] orbital, i.e. on the ground state.     In Fig.~\ref{Fig:espectra} we can see that, as for the other excited bands, this band lies somewhat higher than the experimental one.   The wave function  of the band head is plotted in panel d) of Fig.~\ref{Fig:WFS}.  From its extension and shape, centered around  $\gamma=27^{\circ}$, it looks like a $n_{\gamma}=0$,  $(K+2)$ $\gamma$-band.  The decay of this band to other states is displayed in Fig.~\ref{Fig:band_93}.  In the theoretical results, shown in panel a),  the band head decays mainly  via an E2 transition to the ground state. Experimentally, panel b),  this is also the main branch though it decays also strongly to the $7/2^+$ state at variance to the theory. 
   The $11/2^+$ level decays via E2 to the band head and to the ground band.  This is in agreement with the experimental findings indicating clearly that the $9/2^+$ band is a $(K+2)_{\gamma}$ vibration on the [202 5/2] orbital in agreement with earlier assignments.
    
      The spectroscopic quadrupole moment of the 9/2$^{+}$  state, see Table~\ref{tab:promedio},  is 18  e fm$^2$  in the shell model approach and 30 e fm$^2$ in the rotational model approximation.  The RM value is in a much better agreement with our result of 35.5 e fm$^2$.

  \subsection{The 3/2$^+$ excited band (band V)}  
    The band head of the fourth positive-parity  excited  band (band V)  is an $I=3/2^{+}$ state, again a rather pure $K=3/2$ state, see Table~\ref{tab:pesoK}, and with  large average values of  $\beta=0.699$ and $\gamma= 22.7^\circ$, see Table~\ref{tab:promedio}. The $K=3/2$ value can stem either from an orbital with $\Omega=3/2$, as the result of putting a particle in the   [202 3/2] orbital or making a hole in the [211 3/2], see  Fig.~\ref{Fig:SPE}, or from the  coupling  of some collective $K$ (from the rotor in the particle plus rotor model) to a  $\Omega=1/2$ orbital, like the [211 1/2]. In general will be a  combination of both.

      The wave function of the band head, plotted in panel e) of Fig.~\ref{Fig:WFS},  is very extended and  rather soft in the oblate direction. A reason for that, is (as we can see in Fig.~\ref{Fig:SPE})  that  on the oblate side the orbital [200 1/2] goes up and the [202 3/2] down, thus favoring the occupation of the latter. 
      The shape of the wave function indicates a collective character and looks like a $\gamma$ vibration.   We find  connecting transitions to the ground state and  to several states  of band II, which  suggest a coupling to a  $|K-2|_{\gamma}$ vibration on band I and/or a coupling of the [211 3/2] orbital to a $(K-2)$ vibration on the  [211 1/2] orbital.  The latter assignment has also been made by Headly et al.
\cite{Headly.88} to the $I=3/2^+$ state at 4360 keV excitation energy.  With respect to many aspects, like the quadrupole moment and  the transition probabilities to band II,  this level shows similarities to the band head of band V. However, in our calculations we find a strong transition to the ground state which has not been observed experimentally.

 \subsection{The 1/2$^-$ negative parity band (band VI)}
   
   In the negative parity channel, the first excited state is obtained by promoting the odd particle to the [330 1/2] orbital. 
   The states of the band are very pure $K=1/2$.  The band has  as expected a large average deformation of $\beta=0.779$, since the first negative parity orbital crosses the Fermi surface at a very large $\beta$-value.  This large value  explains the $K=1/2$ purity.  Experimentally and in our calculations the band head of the lowest band of negative parity is a $3/2^-$ state indicating a  large value of the decoupling parameter \cite{BM.75}.
   The degree of agreement between theory and experiment is very good, see Fig.~\ref{Fig:espectra}, specially for the three lowest members of the band. The wave function of the band head is provided in panel f) of Fig.~\ref{Fig:WFS}.  Since this band is the lowest one of negative parity the maximum of the collective wave function coincides with the minimum of the potential energy surface shown in Fig.~\ref{Fig:PES_NEG}.

\vspace{1cm}

To conclude this section we would like to remark that as mentioned above in the present approach at each point of the $(\beta,\gamma)$ plane we only consider one quasiparticle state in the SCCM ansatz Eq.~(\ref{GCM_ANSA}), namely the lowest one. In principle one could add  excited one (and  three) quasiparticle  states  in Eq.~(\ref{GCM_ANSA}) as it has been done in Refs.~\cite{CE.16,CE.17} for even-even nuclei. 
This generalization will improve the single-particle degrees of freedom but influence very little the collective ones. The pertinent question is whether  this generalization will modify considerably the present results.  Of course the definitive answer 
can only be given once the calculations have been performed. However, with the information that we now have at hand, i.e. from even-even nuclei \cite{CE.16,CE.17}, we
can conclude that the character of the collective bands will not change dramatically. The main effect of the generalization will be to lower the energies of the excited bands. Obviously new rotational bands built on single particle states may appear at higher energies. 

As mentioned in Sect.~\ref{Sect:spe_PES}  the present calculations have been performed with 8 harmonic oscillator shells.  To reach absolute convergence of energy values the consideration of larger spaces will be required. With respect to the relative energies,
we think that the negative parity states will be more sensitive to  the size of the configuration space since the energy minima are located at larger deformations, see Fig.~\ref{Fig:NeutPlusMinus}.  We have calculated these potential energy surfaces with 10 shells and for deformations $\beta \le 1.2$ and the contour lines look similar in both calculations. However, for larger $\beta$ values, the PES softens faster with 10 than with 8 shells.  For the positive parity states the relevant contours remain unchanged. We do not expect that the enlargement of the configuration space will significantly affect the properties of the levels belonging to bands I, II and VI. A larger effect, however, is to be expected for the higher lying bands, i.e., the bands III, IV and V.

\section{Conclusion and outlook }

In conclusion, we have presented  the extension of the SCCM approach, which has  been very successful  in the description of excited states in even-even nuclei and  of ground state properties of odd-even nuclei in the past, to the spectroscopy of odd-A nuclei.  Our approach includes exact blocking  with  conservation of angular momentum and  particle number as well as the fluctuations in the deformation parameters $(\beta,\gamma)$.  In the numerical application we have used the finite range density dependent Gogny force which is well known to  properly reproduce bulk properties all over the nuclear chart. 

We have applied this theory to the description of excited states in $^{25}$Mg. We find six rotational bands of which five are  clearly identified with the experimental counterparts. The energies of the low-lying bands of positive parity (bands I and II) are in very good agreement with the experimental data. The excitation energies of bands III and IV are somewhat higher than the experimental ones.  We find a fifth collective band, band V,  which has no obvious experimental counterpart.  We also find a negative parity band (band VI) whose energies  agree very well with the experimental ones.    The transition probabilities and spectroscopic quadrupole moments in general agree well with the experimental adopted values.

The results for odd-even nuclei with the SCCM theory follow very closely the guidelines of the even-even ones.  That means that for a precise description of the highly-excited bands, either the cranking frequency must be considered as an additional generator coordinate or more one-quasiparticle states (and possibly three-quasiparticle states)  should be included  in the calculation.

\section*{Acknowledgements}
 This work was supported by the Spanish Ministerio de Econom\'ia y Competitividad under contracts FPA2011-29854-C04-04 and FPA2014-57196-C5-2-P.

%
%
%

\end{document}